\documentclass[a4paper]{article}
\usepackage[T2A,T1]{fontenc}
\usepackage[utf8]{inputenc}

\usepackage{apacite}
\usepackage{graphicx}

\usepackage{url} 

\begin{document}

%
%


\title{Global diagnostics of ionospheric absorption during X-ray solar flares based on 8-20MHz noise measured by over-the-horizon radars.}

%
%




\author{Berngardt O.I.
\and
	Ruohoniemi J.M.
\and
	St-Maurice J.-P.
\and
	Marchaudon A.
\and
	Kosch M.J.
\and
	Yukimatu A.S.
\and
	Nishitani N.
\and
	Shepherd S.G.
\and
	Marcucci M.F.
\and
	Hu H.
\and
	Nagatsuma T.
\and
	Lester M.
	}

\maketitle

\begin{abstract}
An analysis of noise attenuation during eighty solar flares between 
2013 and 2017 was carried out at frequencies 8-20 MHz using thirty-four 
SuperDARN radars and the EKB ISTP SB RAS radar.  
The attenuation was determined on the basis of noise measurements
performed by the radars during the intervals between transmitting periods.
The location of the primary contributing ground sources of noise was found
by consideration of the propagation paths of radar backscatter from the
ground.
The elevation angle for the ground echoes 
was determined through a new empirical model. It was used to determine the paths 
of the noise and the location of its source. The method was particularly well suited 
for daytime situations which had to be limited for the  most part to only two crossings 
through the D region.  Knowing the radio path was used to determine an equivalent vertical 
propagation attenuation factor.   The change in the noise during solar flares was correlated 
with solar radiation lines measured by GOES/XRS, GOES/EUVS, SDO/AIA, SDO/EVE, SOHO/SEM and 
PROBA2/LYRA instruments. Radiation in the 1 to 8$\mathring{A}$ and and near 100$\mathring{A}$ 
are shown to be primarily responsible for the increase in the radionoise absorption, and by 
inference, for an increase in the D and E region density. The data are also shown to be consistent 
with a radar frequency dependence having a power law with an exponent of -1.6. This study shows that 
a new dataset can be made available to study D and E region.
\end{abstract}

%
%
%


%
%

%


%
%
%
%

\section{Introduction}

The monitoring of ionospheric absorption at High Frequency (HF), particularly at high latitudes,  
makes it feasible to predict radio wave absoption at long distances and therefore on global scales
\cite{DRAP,DRAP2}. This in turn makes it a useful tool for study of the dynamics of the D and E regions.
 Traditionally, there are several techniques in use
\cite{Davies_1969,Hunsucker_2002}, including constant power
2-6~MHz transmitters (URSI A1 and A3 methods, see for example \cite{Sauer_2008,Schumer2010}),
riometry using cosmic radio space sources at 30-50~MHz (URSI A2 method \cite{Hargreaves_2010}) and
imaging riometry \cite{Detrick_1990}. 
Recently, a large, spatially distributed network of riometers has been deployed to monitor absorption 
\cite{Rogers_2015}. The development of new techniques for studying absorption 
with wide spatial coverage would be valuable for the validation of global ionospheric models and for global absorption forecasting.

A wide network of radio instruments in the HF frequency range is available with the SuperDARN (Super Dual Auroral Radar 
Network \cite{Greenwald_1995,Chisham_2007}) radars and radars close to them in terms of design and software \cite{Berngardt_2015}. 
The main task of the SuperDARN network is to measure ionospheric convection.
Currently this network is expanding from polar latitudes to mid-latitudes \cite{Baker_2007,Ribeiro_2012} and possibly to equatorial latitudes \cite{Lawal_2018}. 
Regular radar operation with  high spatial and temporal resolutions and a wide field-of-view makes them a useful
tool for monitoring ionospheric absorption on global scales. The frequency range used by the radars fills a gap 
between the riometric measurements at 30-50~MHz (URSI A2 method) and radar measurements at 2-6~MHz band 
(URSI A1, A3 methods). Various methods are being developed for using these radars 
to study radiowave absorption. One approach is to monitor third-party transmitters \cite{Squibb_2015} and another is to use the signal 
backscattered from the ground \cite{Watanabe_2014, Chakraborty_2018, Fiori_2018}. In this paper, another method 
is investigated. It is based on studying the attenuation of HF noise in the area surrounding the radar that is measured without transmitting any sounding pulses.

Every several seconds, before transmitting at the operating frequency, the radar measures 
the spectrum of the background noise in a 300-500~kHz band centered on a planned operating frequency that lies between 8-20~MHz. 
The minimum in the spectral intensity is recorded and defined here as the 'minimal HF noise level'.

\cite{BERNGARDT_20181} showed that the dynamics 
of the minimal HF noise level 
is strongly influenced by X-ray 1-8$\mathring{A}$ solar radiation in the daytime. 
This effect has also been observed during solar proton events \cite{Bland_2018} where it was found to
correlate well with riometer observations.
This allows one to use 
the noise measured with HF radars to investigate the absorption processes in the lower part of the ionosphere 
in passive mode, without the use of third-party transmitters, 
and without relying on the presence of backscatter from the ground.

To use this new technique on a regular basis for monitoring ionospheric
absorption we should investigate the observed noise level variations
during X-ray flares and show that the observed dynamics are consistent with 
the current absorption models.

\begin{figure}
\includegraphics[scale=0.5]{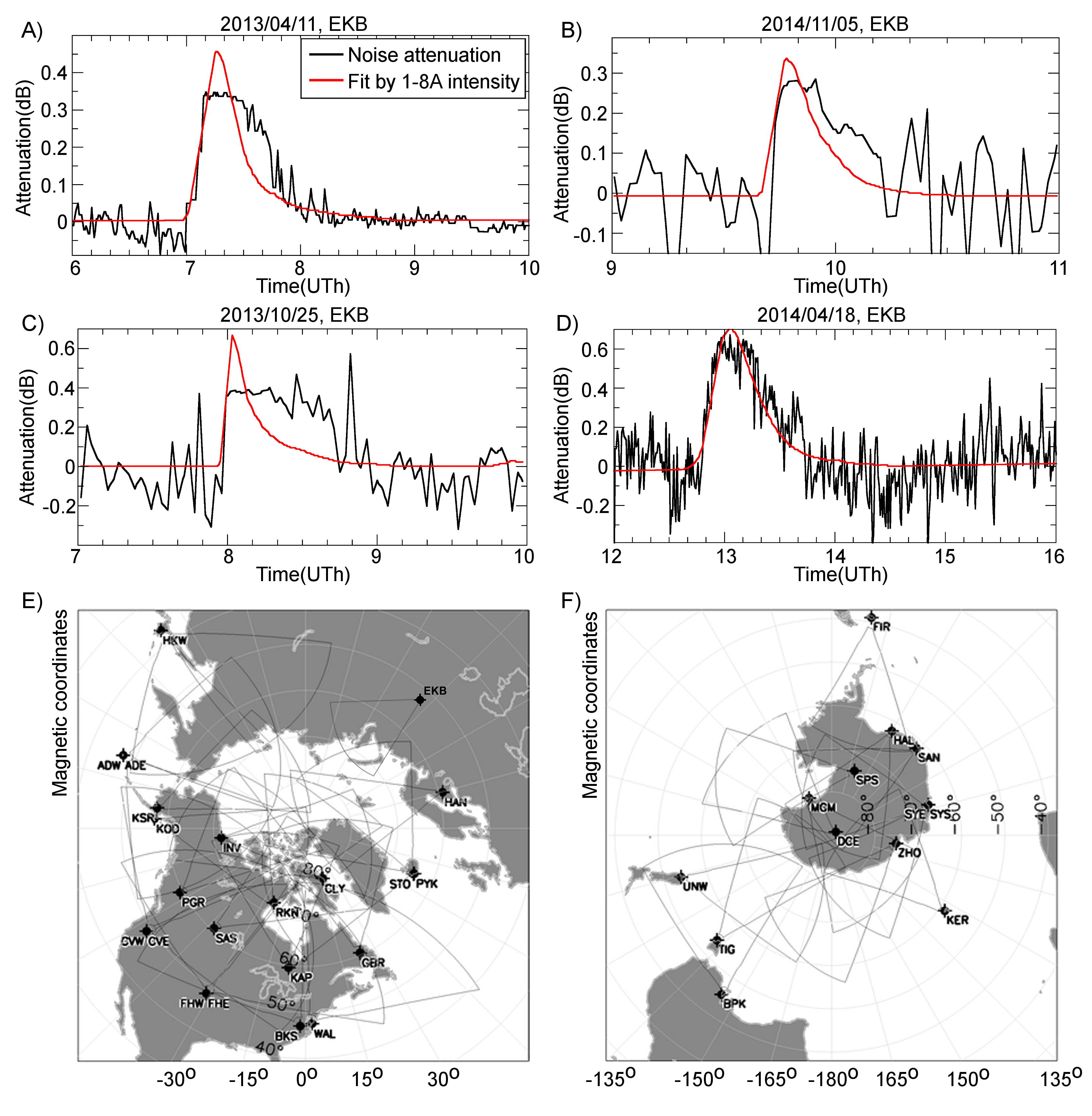}
\caption{A-D) comparison of the X-ray intensity dynamics measured on GOES/XRS 1-8$\mathring{A}$
and the noise attenuation at EKB ISTP SB RAS radar during four flares;
E-F) - fields of views of radars that participated in the work
}
\label{fig:rad_map}
\end{figure}

As shown in the preliminary analysis \cite{BERNGARDT_20181}, 
there is significant correlation of noise level attenuation
with the intensity of X-ray solar radiation in
the range 1-8$\mathring{A}$. However, the temporal
dynamics of the absorption sometimes do not precisely track the solar radiation
at  wavelengths of 1-8$\mathring{A}$, which indicates the presence
of mechanisms other than the ionization of the D-layer
by 1-8$\mathring{A}$ solar radiation. An example of such a comparison will be presented here in 
Fig.\ref{fig:rad_map}A-D and was shown by \cite[fig.9]{BERNGARDT_20181}.

In contrast to riometers which measure ionospheric absorption at relatively
high frequencies (30-50~MHz), the SuperDARN coherent radars use lower operating
frequencies and ionospheric refraction significantly affects
the absorption level - the trajectory of propagation is distorted
by the background ionosphere. To compare the data of different
radars during different solar flares, our method requires taking
into account the state of the background ionosphere during each experiment.
This allows an oblique absorption measurement to be converted to an equivalent vertical
measurement. In addition, the solution of this problem allows determination of
the geographic location of the region in which the absorption takes
place.

Among the factors that affect the error in estimating the absorption
level is the frequency at which the radar operates and its irregular
switching. It is known that the absorption of radio waves depends on frequency, 
but this dependence is taken into account in different ways in different papers. 
In order to make a reliable
comparison of data collected from radars operating at different frequencies, it is necessary to find
the frequency dependence of the HF noise absorption, and to take it into account.
This will allow us to infer the absorption at any frequency from the 
observed absorption at the radar operating frequency.

The third factor that needs to be taken into account is the altitude
localization of the absorption.

The present paper is devoted to solving these problems. 
An analysis is made of 80 X-ray solar flares during the years 2013-2017 ,
which were also considered in \cite{BERNGARDT_20181} based on the available data
of 34 high- and mid-latitude radars of SuperDARN network and on the
EKB ISTP SB RAS \cite{Berngardt_2015} radar data. The radar locations and their fields of view
are shown in fig.\ref{fig:rad_map}E-F, the radar coordinates are
given in the Table \ref{tab:rad_list}.
The X-ray solar flares dates
are listed in \cite{BERNGARDT_20181}.

\section{Taking into account the background ionosphere}

As was shown in \cite{BERNGARDT_20181}, during solar X-ray flares
attenuation of the minimal noise level in the frequency 
range 8-20~MHz is observed on the dayside by midlatitude coherent radars. The attenuation
correlates with the increase of X-ray solar radiation 1-8$\mathring{A}$ and  
is associated with the absorption of the radio signal in the lower
part of the ionosphere. 

The HF radio noise intensity is known to vary with local time due different sources \cite{ITU_R}. 
At night, the noise is mostly atmospheric, and is formed by long-range propagation from noise sources around the world, mostly from regions of thunderstorm activity. 
In the daytime the atmospheric noise level significantly decreases due to regular absorption in the lower part of the ionosphere and the increasing number of propagation hops 
(caused by increasing electron density and lowering of the radiowave reflection point).
As a result, in the daytime the multihop propagation part of the noise becomes small, 
and only noise sources from the first propagation hop (mostly anthropogenic noise) need to be taken into account \cite{BERNGARDT_20181}.

An important
issue related to the interpretation of the noise level is the spatial
localization of the effect. It can be estimated by taking into account the radiowave trajectory
along which most of the noise is received and absorption is taking place.
We will argue that ionization of the lower ionosphere is small enough and skip distance variability is less pronounced than 
the variations caused by other regular and irregular ionospheric variations.

Let us consider the problem of detecting the noise source from the
data of a HF coherent radar. It is known that the intensity of the
signal transmitted by an isotropic source and propagating in an inhomogeneous
ionosphere substantially depends on the ground distance from the
signal transmitter to receiver. If we consider only waves reflecting
from the ionosphere, then at sounding frequencies above $foF2$ there
is a spatial region where the signal cannot be received - the dead
zone. At the boundary of this dead  zone (skip distance) the signal appears and is
significantly enhanced  compared to other distances \cite{Shearman_1956,Bliokh1988}.

More specifically, consider that, due to refraction, the signal
transmitted by a point source produces a non-uniform distribution of
power $P(x)$ over the range $x$. According to the theory of radio
wave propagation, the distribution of signal power is determined by
the spatial focusing of the radio wave in the ionosphere, and has
a sharp peak at the boundary of the dead zone \cite{Kravtsov_1983}.
According to \cite{Tinin1983} in a plane-layered ionosphere, the
distribution of the power over  range is:

\begin{equation}
P(x)\simeq\frac{1}{\sqrt{\sigma_{x}(s_{m})\bar{x}''(s_{m})}}e^{-\frac{\xi^{2}}{4}}D_{-\frac{1}{2}}(\xi)\label{eq:GS_shape}
\end{equation}
where $D_{-\frac{1}{2}}(\xi)$ is the parabolic cylinder function \cite{Weisstein_2018};
$x_{m}$ - the distance at which the spatial focusing is observed; $\xi=\frac{x_{m}-x}{\sigma_{x}(s_{m})}$
is the normalized range  relative to $x_{m}$; $s_{m}$ is the sine of elevation angle; 
$\sigma_{x}(s_{m})$ is the standard deviation of $x$ over the geometric optical rays
; $\bar{x}''$ is second differential of $x$ with respect to $s_{m}$.

Let us consider this signal after it is scattered by inhomogeneities on the Earth's surface 
and then received by the radar. In the first approximation the power of the signal received 
by the radar will be proportional to the product of (i) the power of the incident signal $P(x)$ 
(related to spatial focusing when propagating from the radar to the Earth's surface);
(ii) the scattering cross-section $\sigma(x)$ (related to inhomogeneities of the Earth's surface);
and (iii) the incident power $P(x)$ (related to the propagation from the Earth's surface to the radar).

This 
signal 
is received as a powerful signal coming from
a small range of distances. When analyzing the data of coherent HF
radars, this signal, associated with the focusing of the radio wave
at the boundary of the dead zone, is referred to as ground scatter (GS)
\cite{Shearman_1956}.

The scattering cross section $\sigma(x)$ essentially depends on the 
angles of incidence and reflection of the wave, as well as on the 
properties and geometry of the scattering surface. This causes a 
significant dependence of the GS signal on the landscape and the 
season \cite{Ponomarenko_2010}.  In the case of presence of significant inhomogeneities, 
for example, mountains \cite{Uryadov_2018}, $\sigma(x)$ may cause the appearance of 
additional maxima and minima in the GS signal. For relatively homogeneous surfaces, 
the position of the GS maximum remains almost unchanged, and the GS signal 
propagation trajectory (radar-surface-radar) can be used to 
estimate the trajectory of the propagation of the noise signal (surface-radar). 
Below we use this approximation to localize noise source using GS signal properties.

Let the independent noise sources be distributed over
the Earth's surface over the distance $x$ of the first hop (from 0 to 3000km). Let their intensity be 
$B(x)$ and the radiation pattern 
of each of them be nearly isotropic
 over the elevation angles forming
the GS signal. Let the noise signals interfere incoherently. In this case the power of
the signal $P_{0}(x_{1})$, received at the point $x=x_{1}$, in the
first approximation becomes:

\begin{equation}
P_{0}(x_{1})\simeq\int_{-\infty}^{\infty}B(x)P(x_{1}-x)dx
\label{eq:noise_location}
\end{equation}

\begin{figure}
\includegraphics[scale=0.38]{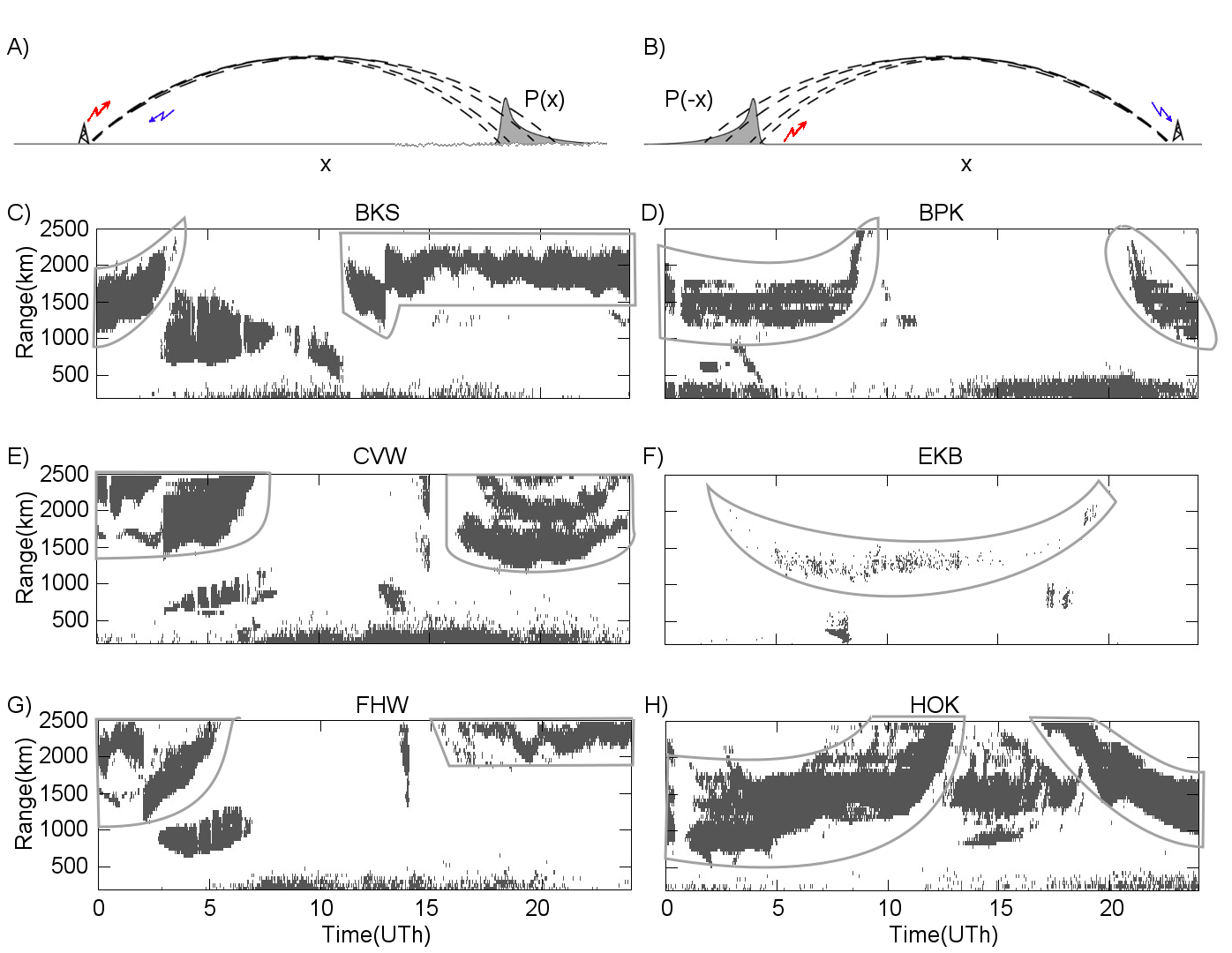}
\caption{
A) - formation of GS signal;
B) - formation of noise power level by distribution of noise sources. Red and blue arrows in A-B) mark transmitted and received signals;
C-H) - the position of the signals, defined by FitACF algorithm as GS, during
18/04/2016 on the radars BKS, BPK, CVW, EKB, FHW, HOK. 
Gray enclosed areas correspond to GS when focusing in the F-layer. 
Other areas are defined by the algorithm, as GS, but having, sometimes, an ionospheric origin.}
\label{fig:geometry_1}
\label{fig:GS_examples}
\end{figure}

Thus, one can represent the formation of the noise power
from terrestrial sources, as a weighted sum of the contributions
from individual noise sources. The function $P(x)$ is the weight,
and the region of localization of the noise source is of the order
of the maximal width of the GS signal (see equation \ref{eq:GS_shape}). According
to the experimental  data it is of the order
of several hundred kilometers.
For the validity of equation (\ref{eq:noise_location}), the characteristic scale of the
homogeneity of the ionosphere in the horizontal direction should be
about the width of the GS signal maximum.
The process of forming the received signal is illustrated in Fig.\ref{fig:geometry_1}B.

Thus, the problem of localization of the noise source can be reduced
to determining the geographic location of the region forming the GS
signal and determining the propagation path of the signal from this
region to the receiver.

In radar techniques, there are a number of procedures for separating the GS signal
from other scattered signal types \cite{Baker_1988, Barthes_1998, Blanchard_2009, Ribeiro_2011, Liu_2012}, but using them for 
automatic location of the effective noise source causes some
problems. To begin with the GS signal can have several ranges
at one time (for example first-hop GS and second-hop GS, or multimode propagation due to mid-scale irregularities \cite{Stocker_2000}). 
It may be discontinuous in time due to defocusing (refraction) and
absorption processes. Finally, it may have irregular temporal
dynamics due to large scale ionospheric variations (for example, internal
atmospheric waves \cite{Oinats2016, Stocker_2000}). 
These problems significantly complicate the automatic interpretation
of the radar data for our task, especially for high-latitude radars where the
ionosphere is essentially heterogeneous with latitude. Therefore, for
automatic estimation of the effective noise location, it was decided to
use a smooth adaptive model of GS position, automatically corrected
by the experimental data.

The study of absorption on the long paths using
GS signal or noise requires knowledge of the trajectory of radio space signal
propagation especially in the two regions where it intersects the D-layer
- near the receiver (radar) and near the transmitter source (point
of focusing, where the GS signal is formed). According to the Breit-Tuve
principle \cite{Davies_1969}, it is sufficient to know the angle of arrival of the GS
signal and the radar range. In practice, however, there are two significant
problems: the separation of the GS signal from the ionospheric scatter
(IS) signal \cite{Blanchard_2009, Ribeiro_2011} and the calibration of the arrival angle measurements
\cite{Ponomarenko_2015, Shepherd_2017, Chisham_2018}.

Fig.\ref{fig:GS_examples}C-H presents examples of the location of signals detected
as GS by the standard FitACF algorithm (used on these radars for signal
processing). It can be seen from the figure that the scattered signal can
include several propagation paths (Fig.\ref{fig:GS_examples}E, 16-24UT), variations
in the GS signal range (associated, for example, with the propagation
of internal atmospheric waves \cite{Stocker_2000,Oinats2016} (Fig.\ref{fig:GS_examples}C, 14-18UT ; Fig.\ref{fig:GS_examples}G,
18-21UT)), as well as ionospheric and meteor trail scattering ( Fig.\ref{fig:GS_examples}C-H, ranges below 400km)\cite{Hall_1997,Yukimatu_2002,Ponomarenko_2016}.
The signal that qualitatively corresponds to  F-layer GS 
is marked at Fig.\ref{fig:GS_examples}C-H by enclosed regions (the modeling results demonstrating this will be shown later
in the paper). These examples demonstrate that the problem of stable
and automatic selection of the GS region associated with reflection
from the F-layer is rather complicated even with use of the standard processing
techniques.

In this study, the position of the F-layer GS signal
was solved for each radar beam separately
and independently. To generate input data for the GS positioning algorithm
for each moment we identify the ranges where the signals have
the maximum amplitude in the radar data. For this purpose we  select only signals
determined by the standard FitACF algorithm to be GS signal. 

Using these prepared input data, we determine the smooth curve of the distribution of GS with range, 
within the framework
of an empirical ionospheric model with a small number of parameters, adapted
to the experimental data. The problem of determining the position
of the GS signal causes certain difficulties connected to the presence
of a large number of possible focusing points associated with the
heterogeneity of the ionosphere along the signal propagation path \cite{Stocker_2000} 
and ionospheric scattered signals incorrectly identified as GS signals. 

For an approximate single-valued solution of this problem, we reformulate
the problem as the problem of producing a GS signal in a plane-layered
ionosphere with a parabolic layer with parameters estimated
from the GS signal. In the framework of the plane-layered ionosphere with
a parabolic F-layer, we have the following expression for the radar
range to the boundary of the dead zone \cite{Chernov1971}:

\begin{equation}
R_{model}=\frac{f_{0}}{f_{oF2}}\left\{ 2h_{mF2}\sqrt{\chi}+\Delta h\cdot ln\left(\frac{1+\sqrt{\chi}}{1-\sqrt{\chi}}\right)\right\} 
\label{eq:model_range}
\end{equation}
where $\chi=\frac{h_{mF2}-\Delta h}{h_{mF2}}$; $h_{min}=h_{mF2}-\Delta h$
- is the minimal height of the ionosphere, obtained from the condition
$N_{e}(h_{min})=0$; $h_{mF2}$ is the height of the electron density
maximum in the ionosphere, obtained from the condition $N_{e}(h_{mF2})=max$;
$f_{oF2}$ is the plasma frequency of the F2 layer; $f_{0}$ is the
carrier frequency of the sounding signal.

In this model, the geometric distance $D$ over the Earth surface
to the point of focusing is defined as \cite{Chernov1971}:

\begin{equation}
D_{model}=R_{model}cos(\Theta_{model})
\label{eq:range2distance}
\end{equation}

The elevation angle $\Theta_{model}$ of the signal arriving from the dead
zone boundary according to this model is calculated as:

\begin{equation}
cos(\Theta_{model})=\sqrt{1-\chi\left(\frac{f_{0}}{f_{oF2}}\right)^{-2}}
\label{eq:model_elevation}
\end{equation}

For interpretation of absorption the elevation angle is very important:
in the model of the plane-layered ionosphere it also corresponds
to the elevation angle in the D-layer, and relates the observed
absorption to absorption of vertically propagating radio space signal.
So, this angle is important for the interpretation of absorption, both
in the case of observing GS \cite{Watanabe_2014, Chakraborty_2018, Fiori_2018} and in the case
of minimal noise analysis \cite{BERNGARDT_20181, Bland_2018}. Most of the radars
do measure the elevation angle. However, since many antenna characteristics in the HF range vary with time
it is very
important to calibrate the angle. This should be performed on each radar
separately and regularly \cite{Ponomarenko_2015, Chisham_2018, Shepherd_2017} 
and requires significant computations. 
To simplify the problem of smooth and continuous calculation of the GS elevation,
we decided to use model calculations of the angle based on propagation in
the adapted ionosphere model. In this sense this method is close to the approach 
used in \cite{Ponomarenko_2015}. 
One needs to just choose a proper ionospheric model.

The reference ionospheric model IRI \cite{Bilitza_2017}
is a median model
and sufficiently smooth in time, but by default it does not correctly describe
fast changes of $foF2$ in some situations, especially at high latitudes \cite{Blagov_2015}. 
This problem becomes especially critical for GS signal range calculations
for sunset and sunrise periods. Searching for one or several IRI parameters
that are constant during the day will not solve the problem, so it
is necessary to use either an adaptive model that more adequately
describes these periods, or to use IRI model corrected for each moment
using data from an ionosonde network \cite{Galkin_2011,Blagov_2015}. 
We use an adaptive model, which is easier to implement 
and does not require additional data and instruments.

The adaptive model of the parabolic-layer ionosphere was used with a nonlinear
model for $foF2(t)$ and constant values for $h_{mF2}$ and $\Delta h$:

\begin{equation}
f_{oF2}(t)=f_{oF2,min}+\left(f_{oF2,max}-f_{oF2,min}\right)\varepsilon(t)
\label{eq:model_fof2}
\end{equation}

\begin{equation}
\varepsilon(t)=\frac{atan\left(\beta\cdot\left(\Theta(t-\Delta T)-\alpha\right)\right)-atan\left(\beta\cdot\left(\Theta_{min}-\alpha\right)\right)}{atan\left(\beta\cdot\left(\Theta_{max}-\alpha\right)\right)-atan\left(\beta\cdot\left(\Theta_{min}-\alpha\right)\right)}
\label{eq:model_fof2-1}
\end{equation}
where $\Theta(t)$ is the cosine of the solar zenith angle at the
radar location as a function of the time $t$; $\Theta_{min},\Theta_{max}$
is the maximal and minimal cosine of the solar zenith angle during
the day; $\alpha,\beta,\Delta T$ are modeled parameters, computed during the fitting procedure. 
More correctly solar zenith angle should be calculated at the point of radiowave absorption 
but in this paper we do not use this. The parameter $\Delta T$ compensates 
the difference in the first approximation.

The required strong nonlinearity of the model during  sunset and
sunrise moments is provided by the $atan()$ function, by the cosine
of the solar zenith angle $\Theta(t)$ and controlled by several parameters:
$\alpha,\beta,\Delta T,f_{oF2,max},f_{oF2,min}$. The model has enough
degrees of freedom to describe the fast dynamics of $f_{oF2}(t)$ during
solar terminator transitions.
Taking into account the diurnal variation of $h_{max},\Delta h$ does not significantly improve the model, 
since these changes can be compensated by changes in the $f_{oF2}$ parameter.

The use of the cosine
of solar zenith angle $\Theta(t)$ and the small time delay $\Delta T$
allows us describing
the GS dynamics during sunrise and sunset more accurately and 
including
 the geographic position of the radar into the
model. The choice of normalization in (\ref{eq:model_fof2-1}) is
made so that $\varepsilon(t)$ takes values in the range [0,1] 
during the day.  Therefore
 $\varepsilon(t)$ reaches its maximal value near noon and its minimal value
near midnight. As a result the model for $f_{oF2}(t)$ (\ref{eq:model_fof2}) also reaches
its maximal value $f_{oF2,max}$ near noon and its minimal value
$f_{oF2,min}$ - near midnight.

When searching for optimal parameters of the model (\ref{eq:model_range}),
the constant height of the maximum $h_{mF2}$ and the half-thickness
of the parabolic layer $\Delta h$ were assumed to be 350~km and 100~km, respectively. The variations allowed in the model are the following:

\begin{equation}
\left\{ \begin{array}{l}
f_{oF2,max}\in[1,33]MHz;\\
f_{oF2,min}\in[\frac{1}{16},\frac{7}{16}]\cdot f_{oF2,max}MHz;\\
\beta\in[1,5];\\
\alpha\in[-1,1];\\
\Delta T\in[0,3]hours
\end{array}\right.
\label{eq:fit_borders}
\end{equation}

An important problem in approximating the experimental data is the
fitting method. A feature of the GS signal is its asymmetric character
(\ref{eq:GS_shape}): it has a shorter front at ranges below GS signal power maximum, and a
longer rear at ranges above GS signal power maximum. Therefore, the distribution of
errors in determining the GS signal can be asymmetric near the mean value. Because of 
this, the use of the standard least squares method, oriented to "white" symmetrical
noise, can produce a regular 
error.  The existence of
ionospheric scattering and several
propagation modes aggravates the situation even more and substantially
increases the approximation errors.

To improve the accuracy of the approximation, a special
 fitting method has been developed 
to detect GS-signal smooth dynamics
in the presence of signals not described by the GS model.
The fitting method consists of three stages. At the
first stage, the preliminary fitting of the model is made. This stage is required
for preliminary rejection of ionospheric scattering and possible additional
modes of propagation. At the second stage, we reject those signals,
which differ significantly by range from the model. At the third stage,
the final fitting of the model is made. During the first and third stages,
a genetic algorithm is used \cite{simon2013}, as a method of searching for an optimum,
but with different input data and with different functionals of the
optimum. At the second stage a kind of cluster analysis \cite{bailey1994} is used.

\begin{figure}
\includegraphics[scale=0.45]{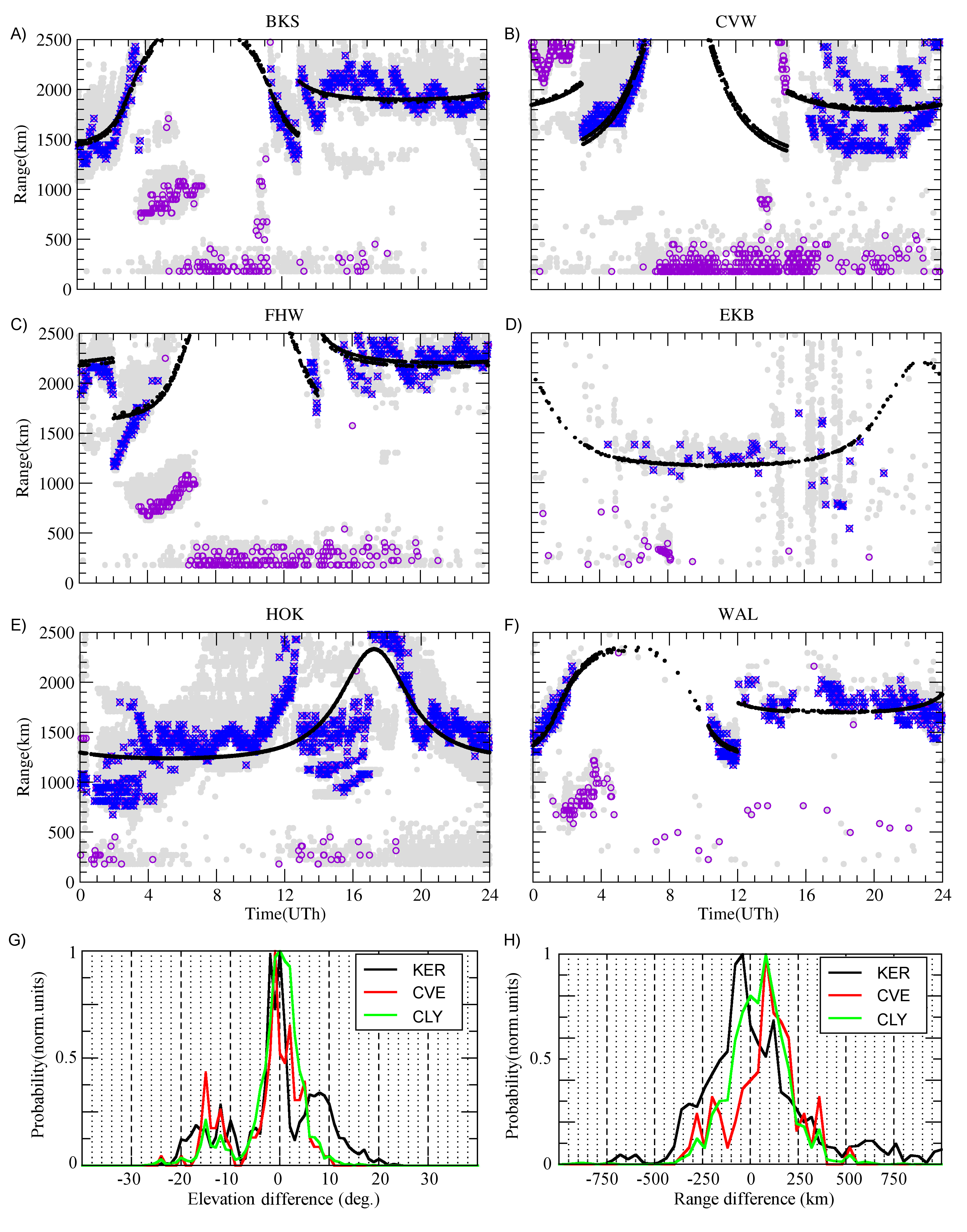}
\caption{A-F) Illustration of the work of the fitting technique on various radars during 18/04/2016.
Violet - non-GS data, detected at the second stage;
blue - GS data, used for 3rd stage; 
black - GS distance, detected at 3rd stage.
G) - the distribution of difference between model and measured GS elevation angles
according to the KER, CVE and CLY radar data 18/04/2016.
H) - the distribution of difference between model and measured GS range
according to  KER, CVE and CLY radar data 18/04/2016.
}
\label{fig:all_rads}
\end{figure}

An illustration of the algorithm operation is shown in Fig.\ref{fig:all_rads}A-F
for 18/04/2016 experimental data. 
There is
good correspondence between the model range and the regular dynamics
of the power of the scattered signal, which indicates a generally
good stability of the technique. Violet circles denote the points
of the GS, extracted from the radar data and serve as input for the first algorithm stage.
The blue crosses denote the points that passed the second
stage (exclusion of ionospheric scattering). The black lines represent the
model dynamics of the GS signal range calculated at the third stage. The line can be discontinuous due to changes 
of radar operational frequency or night propagation conditions.
It can be seen from the figure that qualitatively the technique fits
the GS radar range quite well.

Let us describe the fitting stages in detail.

The points participating in the first stage fitting were determined
by the following condition:

\begin{equation}
R_{exp}(Bm,t)=argmax_{R}(P(Bm,t,R):GSFLAG(Bm,t,R)=true)
\label{eq:gs_detect}
\end{equation}
where $Bm$ is the beam number, $t$ is the time, $GSFLAG$ is the GS
attribute at the given range, calculated by the standard FitACF algorithm \cite{Ponomarenko_2006}
.
The selection rule (\ref{eq:gs_detect}) means that at each moment and on each beam a single
point is found in which the power of the scattered signal is
the maximal over all the signals defined as a GS at this moment and
this beam. Thus, 
at each moment and for each beam, not more than a single point is selected, which
 is used later for fitting. A complete set of points participating
in the fitting on a single beam is shown in Fig.\ref{fig:all_rads}A-F by violet circles.

At the first stage, the fitting of the model (\ref{eq:model_range},\ref{eq:model_fof2},\ref{eq:fit_borders}) 
is made over these selected points (this corresponds to 24 hours of measurements at a single beam).
In order to reduce the error in the presence of ionospheric scatter and additional modes, 
we used 
the following optimizing condition for the fitting:

\begin{equation}
\Omega(Bm)=\sum_{i=0}^{N}W(\delta R_{exp,i})=max
\label{eq:sigma-1}
\end{equation}
where $N$ is the total number of selected points (\ref{eq:gs_detect}) in the data involved in the
fitting, and $W(\delta R_{exp,i})$ is the weight function. The maximization
function (\ref{eq:sigma-1}) and the determination of the
ionospheric parameters are carried out separately for each beam $Bm$.
We do not require these model parameters to be close to each other on different beams.
Our aim is to get smooth and 
physically reasonable
radar distances and elevation angles.  Their
 correctness will be discussed later.

The difference $\delta R_{exp,i}$ of the experimental range from
the model range is defined as:

\begin{equation}
\delta R_{exp,i}=R_{model,i}-R_{exp,i}
\end{equation}

Due to the asymmetric structure of GS signal over range, 
an asymmetric weight function $W$ was chosen:

\begin{equation}
W(\delta R_{exp})=\left\{ \begin{array}{l}
e^{-\frac{\delta R_{exp}}{200[km]}};\delta R_{exp} \geq 0\\
e^{\frac{\delta R_{exp}}{20[km]}}; \delta R_{exp} <0
\end{array}\right.
\end{equation}

This function $W$ takes its maximal value when the experimental data
coincide with the model data ($\delta R_{exp}=0$), and falls to zero if they differ too
much ($|\delta R_{exp}|\rightarrow\infty$).

The choice of characteristic scales of 20 and 200 km is related to
the characteristic durations of the edges of the GS signal. It is obvious
that using such a weight in white noise conditions give a biased
estimate - the model curve passes on average not in the middle of
the experimental points set, but closer to its lower boundary, approximately
with the ratio 1:10. However, in this problem
the result corresponds well to the physical meaning
and structure of the GS signal: its maximal power position is shifted
to smaller distance, so this should qualitatively compensate
the 'non-whiteness' of the observed GS range variations. It should set the model of GS range 
closer to 
reality than the range
calculated by the standard least-squares method. 
On the other hand, the use of such a weight function
makes it possible to minimize the contribution of points substantially
away from the model track (these are ionospheric scatter and other propagational modes) 
and to discard them from consideration during fitting. 

As shown by qualitative analysis, the use of the weight function
makes it possible to increase the stability of the technique in the
presence of other modes and ionospheric scatter, and to carry out a model track
near the lower boundary of the experimental GS data, which corresponds
to the maximal energy of the GS signal.

The second stage of the algorithm is the rejection of ionospheric scattering and other propagation modes 
from the data. It is based on the cluster analysis technique, 
and close to the one used in \cite{Ribeiro_2011}. All the points are put into range-time
 grid of values (100x100). Thus the normalized range and moment of each
point are scaled to integer  values [0,100]. For all the
combinations of such points (i.e. pairs), an Euclidean distance
is calculated, and the points are divided into a clusters based on the distances between them. 
Every point in a single cluster has a nearest neighbor point in the same cluster 
at distance that does not exceed the doubled median distance calculated over the whole dataset. 
This allows us to separate the dataset into isolated clusters.

If the optimal model GS curve, calculated at the first stage, crosses a cluster at least at one point, the whole
cluster is considered a GS signal. Otherwise the cluster is considered
as not GS signal, and all the cluster points are 
excluded from subsequent consideration. The signals defined in the second stage as GS signals
are shown by blue crosses in the Fig.\ref{fig:all_rads}A-F, other
signals are rejected at this stage and marked in the Fig.\ref{fig:all_rads}A-F by violet circles.

In the third stage we believe that only F-layer GS signal points exist in the filtered data, 
and 
we can use the traditional least squares method to fit the model GS
range function to the data:

\begin{equation}
\Omega(Bm)=\sum_{i=0}^{M}\delta R_{exp,i}^{2}=min
\label{eq:sigma-1-1}
\end{equation}
where $M$ is the number of GS points remaining after the second stage. 
The fitting of the modelled GS range at the third stage is
shown in the Fig.\ref{fig:all_rads}A-F by the black line.

In Fig.\ref{fig:all_rads}A-F 
one can also see conditions for which the algorithm does not work well. This happens when ionospheric
 scattering appears
at distances that are close to the daytime GS distance (Fig.\ref{fig:all_rads}E, 00-03UT, 12-17UT;
Fig.\ref{fig:all_rads}F, 15-19UT). Since X-ray solar flares effects are observed mostly during the day \cite{BERNGARDT_20181},
the nighttime areas are not statistically important for this paper. So we do not pay attention to possible nighttime model range errors.
A more critical problem is the case when the 1st and 2nd hop signals
(Fig.\ref{fig:all_rads}B, 17-24UT) are observed equally clearly and with nearly the same amplitude. 
So the model signal is forced to
pass in the middle between these tracks. In this case, a significant
regular error appears.
Therefore, for a small amount of validated data,
(Fig.\ref{fig:all_rads}D), the algorithm can fail.

The model results have been compared with measurements made by the 
polar cap (CLY), sub-auroral (KER) and mid-latitude (CVE) radars on 18/04/2016.
The root-mean-square
error between the model elevation angle and the experimental
measurements calculated from the interferometric data 
is $6-9^{o}$, with an average error of $1-3^{o}$ (Fig.\ref{fig:all_rads}G).
The root-mean-square
error between the model GS range and the experimental
measurements calculated for 18/04/2016 for these radars
is {166-315~km} , with an average error of 7-47~km (Fig.\ref{fig:all_rads}H).
The comparison shows that the technique can be used for processing polar cap, sub-auroral, and mid-latitude radar data.

In conlusion, in most cases, the algorithm works well enough to enable proper statistical conclusions.
The smallness of the average range and elevation angle  errors
make it possible to use this technique for determining the model GS to carry out
statistical studies on a large volume of experimental radar data.

Finally, to identify which hop produces most of the noise absorption, we analyzed
the cases when the 1st hop and 2nd hop GS signal 
locations are at opposite sides of the solar terminator (i.e. in lit and unlit regions). We studied only cases when 
the noise absorption correlates well with X-rays at 1-8$\mathring{A}$. The 2nd hop GS 
distance was estimated by doubling the first hop GS distance
 (\ref{eq:range2distance}). This allows us to estimate geographical location 
of 2nd hop GS region. Since the absorption correlating with X-rays is mainly associated 
with the lit area \cite{BERNGARDT_20181}, the studied 
cases allow us to statistically identify the 
(lit) hop of most effective absorption. 
For the $\approx 400$ cases found with 
 the correlation coefficient $R>0.6$ the 
probability of the absorption at the 1st hop is $78\%$. 
For the $\approx 70$ cases found with $R>0.9$
the probability of absorption at the 1st hop is $95.5\%$.

We made a similar comparison of the point above the radar and the point near the edge of the GS region. Our analysis has shown that
the probability of absorption near GS region for $R>0.8$ (over 15 cases) is $54\%$, 
for $R>0.85$ (over 10 cases) is $75\%$ , and for $R>0.9$ (over 4 cases) is $100\%$.

Therefore, in most situations, the daytime noise absorption can be interpreted as absorption on the 1st 
hop, with the most probable location near the dead zone.

\section{Dependence of the absorption on the sounding frequency}

Using the model of the GS signal range described above, it is possible
to automatically estimate the elevation angle of the incoming noise signal
and, thereby, to transform the oblique absorption to the
vertical absorption. Knowing the height 
of the absorbing region and the range to GS, it is possible 
to estimate the geographical position of the absorbing region. 

Another important factor that needs to be taken into 
account is the frequency
dependence of the absorption.
Using it one can interpolate the absorption measured at the radar operating frequency 
to the absorption at a fixed frequency.
At present, several variants of absorption frequency dependence are
used in the analysis of experimental data and forecasting. The DRAP2
model \cite{DRAP,DRAP2} and some nowcast PCA models \cite{Rogers_2015}  
use a frequency dependence given by $A[dB]=A_{0}f^{-1.5}$, based
on \cite{Sauer_2008}. A frequency dependence $A=A_{0}f^{-1.24}$
is proposed in \cite{Schumer2010}. From the theory of propagation
of radio waves, the frequency dependence for sufficiently high probing
frequencies exceeding the collision frequency $2\pi f\gg\nu$
absorption should have the dependence $A=A_{0}f^{-2}$ \cite{Davies_1969,Hunsucker_2002}.
Computational models like \cite{Eccles_2005,Pederick_2014} 
use an ionospheric and a radio wave propagation model
 to calculate the
absorption on each particular path and do not use 
an explicit frequency dependence.

To perform a comparative statistical analysis on a larger radar dataset, 
it is necessary to retrieve the experimental dependence
of the absorption on the frequency of the radar. To determine this
dependence, a correlation analysis of the absorption at various frequencies
was carried out. 
We selected 'multi-frequency 
experiments', that is, experiments for which, during 6 minutes,  a certain 
radar simultaneously operated at
least on 2 frequencies, separated by at least 10\%, at the same azimuth.
After selecting these 
experiments we built regression coefficients
 between the noise levels at different
frequencies for each 'multi-frequency experiment' , taking into account
the possibility of different background noise levels and their various
(linear) time dependencies. Thus, the regression coefficient $A_{0}$
for each 'multi-frequency experiment' was determined as the value
minimizing the root-mean-square deviation of noise attenuation $P_{1}(t),P_{2}(t)$
at frequencies $f_{1},f_{2}$ 
respectively.  In other words, $A_0$ is defined as the solution to
 the problem:

\begin{equation}
\Omega=\int_{T_{flare}-1h}^{T_{flare}+2h}\left(P_{1}(t)[dB]-\left\{ A_{0}P_{2}(t)[dB]+A_{1}+A_{2}t\right\} \right)^{2}dt=min
\end{equation}

The integration was made over the regions $P_{1}(t)<0.9\cdot max(P_{1}),P_{2}(t)<0.9\cdot max(P_{2})$  
to exclude noise saturation effects from consideration. To increase the validity of the retrieved data, 
we analyzed only the
cases where the correlation coefficient between the noise attenuation and the
variations of the 
intensity of solar radiation in the 1-8$\mathring{A}$ band exceeded
 0.4, which indicates a statistically significant absorption
effect \cite{BERNGARDT_20181}. As a result, we obtained a statistical
distribution of the exponent of the power-law dependence of the absorption
on the frequency 
\begin{equation}
A[dB]\sim f^{-\alpha}
\label{eq:abs_alpha}
\end{equation}
by calculating the ratio for every experiment:
\begin{equation}
\alpha_{i}=\frac{log(A_{0,i})}{log(f_{1,i}/f_{2,i})}
\label{eq:alpha_rel}
\end{equation}
where $f_{2,i},f_{1,i}$ are the frequencies of noise observation simultaneously
on the same beam at the same radar, and $A_{0,i}$ is the coefficient
of regression between the absorption and X-ray flare dynamics at different
sounding frequencies; $i$ is the experiment number.

Fig.\ref{fig:alpha_distr}A shows the 
parameters of statistical distribution of $\alpha$
calculated over 'multi-frequency experiments'
for relatively high frequency difference ($f_{1}/f_{2}\in[1.2,1.3];f_{1}/f_{2}\in[1.3,1.5];f_{1}/f_{2}\in[1.5,1.6]$)
and absorption for correlating 
($|R|>0.4$) with 1-8$\mathring{A}$ solar radiation. 
To improve the
estimates, we selected only
experiments with 
small carrier frequency variations $\delta f_{1},\delta f_{2}$ during flare observations 
($|\delta f_{1}|,|\delta f_{2}|<150kHz$) around the average sounding frequencies ($f_{1}, f_{2}$).
In other words, we investigated multi-frequency experiments with a large enough 
difference between two frequencies, that is, we required

\begin{equation}
| f_{1} - f_{2} | > 3 \cdot (| \delta f_{1}| + | \delta f_{2} |)
\label{eq:select_cond}
\end{equation}

This final distribution corresponds to 1662 individual experiments at 18 different
radars (BKS, BPK, CLY, DCE, EKB, GBR, HKW, HOK, INV, KAP, KOD, KSR, MCM, PGR, RKN, SAS, TIG, WAL). 
It can be seen from Figure \ref{fig:alpha_distr} that the distribution of $\alpha$ has an average 
around 1.6 (for $f_{1}/f_{2} > 1.3$)
and RMS can reach about 0.3 (at $f_{1}/f_{2} > 1.5$). 
The statistics indicate that the
 dependence of the absorption on the frequency
in the range 8-20 MHz can be described more stably by the empirical
dependence $A[dB]\sim f^{-1.6}$, which is close to $\alpha=1.5$,
used in the conventional absorption forecast model DRAP2 \cite{DRAP,DRAP2}.
Therefore, we will use the empirically found value $\alpha=1.6 \pm 0.3$ 
in the following work.

\section{Correlation of absorption dynamics with solar radiation of different
wavelengths}

The next important issue arising in the investigation of noise data
by coherent radars is the interpretation of the detailed temporal
dynamics of the noise absorption. As shown in \cite{BERNGARDT_20181}
and seen in fig.\ref{fig:rad_map}A-C, the front
of noise absorption at the radar correlates well with the shape
of the X-ray flare according to GOES/XRS 1-8$\mathring{A}$. 
The rear
is substantially delayed with respect to the X-ray 1-8$\mathring{A}$
flare. As the preliminary analysis showed,
this is a relatively regular occurrence for the data from 2013 to 2017.  
Since the absorption from the rear is delayed for tens of minutes, it cannot 
be explained only in terms of recombination in the ionized region.

One possible 
explanation for the delay in the rear is the contribution to
ionospheric absorption of regions higher than the D layer,
ionized by solar radiation lines other than the
X-ray 1-8 $\mathring{A}$. It is known that the lower part of the
ionosphere (layers D- and E-) is ionized by wavelengths \textless{}100
$\mathring{A}$ \cite{banks1973aeronomy} as well as by Lyman-$\alpha$
line (about 1200$\mathring{A}$). Most often, researchers analyze
the association of absorption with X-ray radiation 1-8 $\mathring{A}$
only, measured by GOES/XRS and associated with the ionization of the
D-layer \cite{Rogers_2015, Warrington_2016}, see fig.\ref{fig:rad_map}D. 
However, the absorption is important not only in the D-layer
but also in the E-layer, the ionization of which is caused by other
components of the solar radiation. In particular, 
soft X-ray 10-50 $\mathring{A}$ radiation  is taken into account
in modern D-layer ionization models \cite{Eccles_2005}
(where it is taken into 
account using a solar
 spectrum model) . The combined effect of increasing absorption
in the E-layer and a slight refraction extending
the path length in the absorbing layer leads to the need to take into
account the ionization of the E-layer.

To analyze the correlation of the noise attenuation with various solar radiation
lines, we carried out a joint analysis of the absorption during the
80 flares of 2013-2017 and 
data from varied  instruments, namely:
 GOES/XRS \cite{Hanser_1996,Machol_2016b},
GOES/EUVS \cite{Machol_2016}, SDO/AIA \cite{Lemen_et_al_2012}, PROBA2/LYRA \cite{Hochedez_2006,Dominique_2013}, SOHO/SEM \cite{Didkovsky_2006}, SDO/EVE(ESP) \cite{Didkovsky_2012}.
These instruments provide direct and regular
 observations of solar radiation
in the wavelength range 1-2500$\mathring{A}$ during the period under
study (see Table \ref{tab:sol_wl} 
for details). 
It is well known that at different wavelengths
the solar radiation dynamics during flares is different
\cite{Donnelly_1976}.
This allows us to find the solar radiation lines that most strongly 
influence
the dynamics of the noise variations at the coherent radars.

To determine the effective ionization lines, we calculate the following probability:

\begin{equation}
P\left({\Lambda}\right)=P\left( R(P(t),I_{\Lambda}(t))\ge R(P(t),I_{1-8\mathring{A}}(t)) | R(P(t),I_{1-8\mathring{A}}(t))\ge 0.4 \right)
\label{eq:P_RR}
\end{equation}
 In this expression, $P\left({\Lambda}\right)$ is the probability that the correlation coefficient $R(P(t),I_{\Lambda}(t))$ 
of the observed absorption $P(t)$ with the intensity $I_{\Lambda}(t)$ 
of a given solar radiation line $\Lambda$ during the X-ray flare period will
not be lower than the correlation coefficient $R(P(t),I_{1-8\mathring{A}}(t))$ of the observed absorption $P(t)$
with the intensity $I_{1-8\mathring{A}}(t)$ of GOES/XRS 1-8$\mathring{A}$ line. 
The calculations 
are carried out only for cases during which the correlation
coefficient between absorption and GOES/XRS solar radiation is greater
than 0.4. 

\begin{figure}
\includegraphics[scale=0.35]{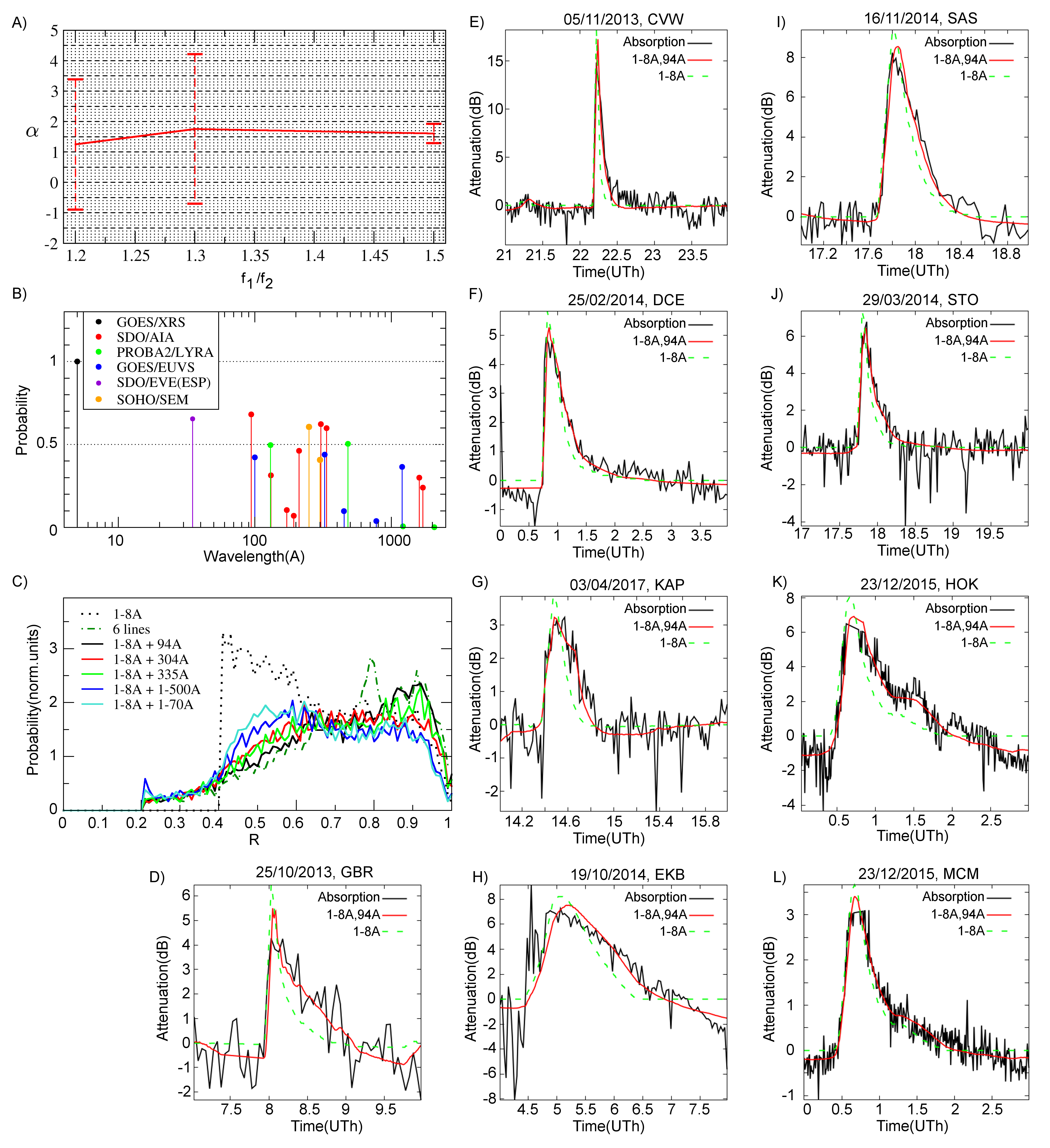}
\caption{
A) Average and RMS of the power-law (\ref{eq:abs_alpha}) coefficient $\alpha$ of the absorption
frequency dependence as a function of relation of frequencies during multi-frequency experiments;
B) The probability $P\left({\Lambda}\right)$ (\ref{eq:P_RR}) over all the flares and the radars;
C) Distribution of correlation coefficients for various approximations
of the noise absorption experimental data; 
D-L) are examples of fitting the
attenuation of HF noise by different combinations of solar spectrum
lines (at different radars during different X-ray flares).
}
\label{fig:alpha_distr}
\label{fig:lines_corrr}
\label{fig:fit_several_lines}
\end{figure}

It should be noted that if the distribution of values of the correlation
coefficients are similar and independent for different wavelengths
of solar radiation, then 
 $P(\Lambda)$ should not exceed 0.5.  Exceeding this level indicates a line of solar radiation 
to be a controlling factor for the attenuation of the noise.  Figure \ref{fig:lines_corrr}B shows 
the results of this analysis based on the processing of over 11977 individual observations.

One can see from Figure \ref{fig:lines_corrr}B that very often (in 62 to 68\% of the cases) $P(\Lambda)$ exceeds 0.5 for $\Lambda$ in the ranges SDO/AIA 94$\mathring{A}$,
SDO/EVE 1-70$\mathring{A}$, 300-340$\mathring{A}$, SDO/AIA 304,335$\mathring{A}$, SOHO/SEM 1-500$\mathring{A}$.  This indicates
the need to take these solar radiation lines
into account when interpolating the HF noise attenuation.
All these lines are absorbed below 150~km \cite[fig.2]{TOBISKA2008803}.  They are therefore sources of ionization  in the lower part of the ionosphere and are
contributing to  the radio noise absorption observed in the experiment.

Let us demonstrate the potential of using the linear combination of
six lines from these spectral ranges (1-8$\mathring{A}$, 94$\mathring{A}$,
304$\mathring{A}$, 335$\mathring{A}$,  1-70$\mathring{A}$,  1-500$\mathring{A}$) instead of just single 1-8$\mathring{A}$
GOES/XRS line. Let us assume that ionization
is produced by different lines independently, the contributions
of each line to ionization are positive, 
and are retrievable.   To search for the amplitude of these contributions
, we used the non-negative
least-squares method \cite{Lawson_Hanson_1995}. It provides an
iterative search for the best approximation of experimental noise
attenuation $P_{att}(t)$ by a linear combination of solar radiation
dynamics at different wavelengths ($P_{1-8\mathring{A}}(t)$, $P_{94\mathring{A}}(t)$, $P_{304\mathring{A}}(t)$, $P_{335\mathring{A}}(t)$, $P_{1-70\mathring{A}}(t)$, $P_{1-500\mathring{A}}(t)$)
with unknown nonnegative weighting multipliers. In addition we also take into account slow background noise dynamics 
by adding a linear dependence $C_{0}+C_{1}t$ into the regression.

Finally, we search for parameters $C_{0..7}$ that solve the problem:

\begin{eqnarray}
 \int_{T_{flare}-1h}^{T_{flare}+2h} (P_{att}(t)-C_{0}-C_{1}t-C_{2}P_{1-8\mathring{A}}(t)-C_{3}P_{94\mathring{A}}(t) -C_{4}P_{304\mathring{A}}(t) \\
-C_{5}P_{335\mathring{A}}(t)-C_{6}P_{1-70\mathring{A}}(t)-C_{7}P_{1-500\mathring{A}}(t))^{2} dt =min
\label{eq:full_att_model}
\end{eqnarray}
under the limitation that $C_{2},C_{3},C_{4},C_{5},C_{6},C_{7}$ be all positive.

Examples of approximations
and statistical results are shown in
Fig.\ref{fig:fit_several_lines}C-F. 
It can be seen that the sum of four lines (dot-dashed green line) approximates the experimental
data 
much better than just a single GOES/XRS (dotted black line) solar radiation
line. 
Fig.\ref{fig:fit_several_lines}C shows the distribution
of the correlation coefficients when the experimental data are approximated
by linear combinations of the lines 1-8$\mathring{A}$, 94$\mathring{A}$,
304$\mathring{A}$, 335$\mathring{A}$, 1-70$\mathring{A}$, and 1-500$\mathring{A}$ . 
The figure shows that
the combination of the lines 1-8$\mathring{A}$ and 94$\mathring{A}$
(solid black line) fits the experimental data no worse than the combination
of all six lines (dot-dashed green line), and significantly better than the single line 1-8$\mathring{A}$
(dotted black line). This allows us to use a combination of the two lines 1-8$\mathring{A}$
and 94$\mathring{A}$ as parameters of the noise attenuation model
during X-ray solar flares at these radars. 

In this paper we analyze only X-ray flares, and the level of Lyman-$\alpha$ line is comparatively weak. 
Therefore the well-known dependence of the D-layer ionization with Lyman-$\alpha$ is not detected (see Fig.\ref{fig:fit_several_lines}B).

Lines 10-100$\mathring{A}$
are usually absorbed at heights of the order of and below 100 km \cite[fig.1.7, par.6.3.]{banks1973aeronomy},
This indicates a significant contribution of the lower part of the
E-layer to the noise absorption observed by the radars.

 The median
value of the correlation coefficient of the noise attenuation with
1-8$\mathring{A}$ is 0.62, with the combination of 1-8$\mathring{A}$
+ 94$\mathring{A}$ lines is 0.76, and with the combination of
all 6 lines is 0.73.

Thus, taking into account the line 94$\mathring{A}$ leads to an increase
in the median correlation coefficient from 0.62 
to 0.76, while adding other lines
does not significantly increase the correlation. This allows us 
concluding
that use of the 1-8$\mathring{A}$
and 94$\mathring{A}$ solar radiation lines as a proxy of the noise
attenuation profile potentially allows a more accurate approximation
of the temporal dynamics of the experimentally observed noise attenuation, 
and as a result, of the temporal
 dynamics of the absorption of the HF radio signals in the
lower part of the ionosphere. 
Fig.\ref{fig:fit_several_lines}D-L shows the attenuation of HF noise dynamics
 when
it is approximated only by GOES/XRS 1-8$\mathring{A}$ (green dashed
line) and by a combination of GOES/XRS 1-8$\mathring{A}$
and SDO/AIA 94$\mathring{A}$ solar radiation (red line). The
approximations are shown for several radars during several flares. It
can be seen from the figure that taking into account intensity of the SDO/AIA 94$\mathring{A}$ line
significantly improve the accuracy of fitting the noise
attenuation dynamics.
 Therefore it is necessary to take into account
not only the D-layer, but also the E-layer of the ionosphere for the interpretation of the noise absorption
during X-ray solar flares. This corresponds well
with the results obtained by \cite{Eccles_2005}.

\section{Diagnostics of global absorption effects}

Taking into account all of the above, it is possible to build an automatic
system suitable for global analysis of ionospheric absorption of HF
radio waves over the area covered by radar field-of-views. The algorithm
for constructing the automatic absorption analysis system consists
of the following stages.

At the first stage, the GS signal range curve is determined from the
daily behavior of the GS 
signal. We model the ionosphere as a parabolic layer of known half-thickness
$\Delta h$ and height $h_{mF2}$, but of unknown amplitude $f_{oF2}(t)$ and dynamics.
The temporal dynamics of $f_{oF2}(t)$ is 
approximated by the nonlinear parametric
 function (\ref{eq:model_fof2}), and 
its parameters are calculated from experimental data 
via a fitting procedure.

Using this GS signal range curve, the elevation angle of the received GS signal
is estimated as a function 
of time.   The location of the region making the main contribution to the absorption of the
radio noise is found simultaneously.
 Its calculation is based on the Breit-Tuve principle
\cite{Davies_1969} and on the
assumption that the signal is reflected at the virtual height
 $h_{mF2}$. Such a calculation is carried out separately
for each radar, for each beam. The algorithm for constructing the
dynamics of GS range and the elevation angle is given
above (\ref{eq:model_range},\ref{eq:model_elevation}).

At the second stage, the noise absorption level $\widetilde{P}_{vert,10MHz}(t,\phi(t),\lambda(t))$
is estimated for the vertical radio wave propagation in the absorbing
layer at a frequency of 10MHz for each beam of the radar, at a geographical
point $(\phi(t),\lambda(t))$ corresponding to the position of the effective absorbing region.
It is calculated from the noise variations $\widetilde{P}(t)$ detected by the radar, taking into account 
the elevation angle $\Theta_{model}$ of the radio signal propagation in
the absorbing layer, which was calculated at the first stage. The
absorption corresponds to the geographic coordinates $(\phi(t),\lambda(t))$, also calculated
in the first stage, and set to the 
point which is farthest away from the radar (the trajectory crosses D-layer at two points). 
The observed dependence of absorption on frequency $f(t)$ is
interpolated to 10MHz 
frequency using our retrieved median frequency dependence. 
The resulting expression for the vertical absorption is:

\begin{equation}
\widetilde{P}_{vert,10MHz}(t,\phi(t),\lambda(t))=\widetilde{P}(t)sin(\Theta_{model}(t))\left(\frac{f(t)}{f_{0}}\right)^{1.6}
\label{eq:final_eq}
\end{equation}
where $f_{0}=10$MHz, and $f(t)$ is the radar sounding frequency.

\begin{figure}
\includegraphics[scale=0.15]{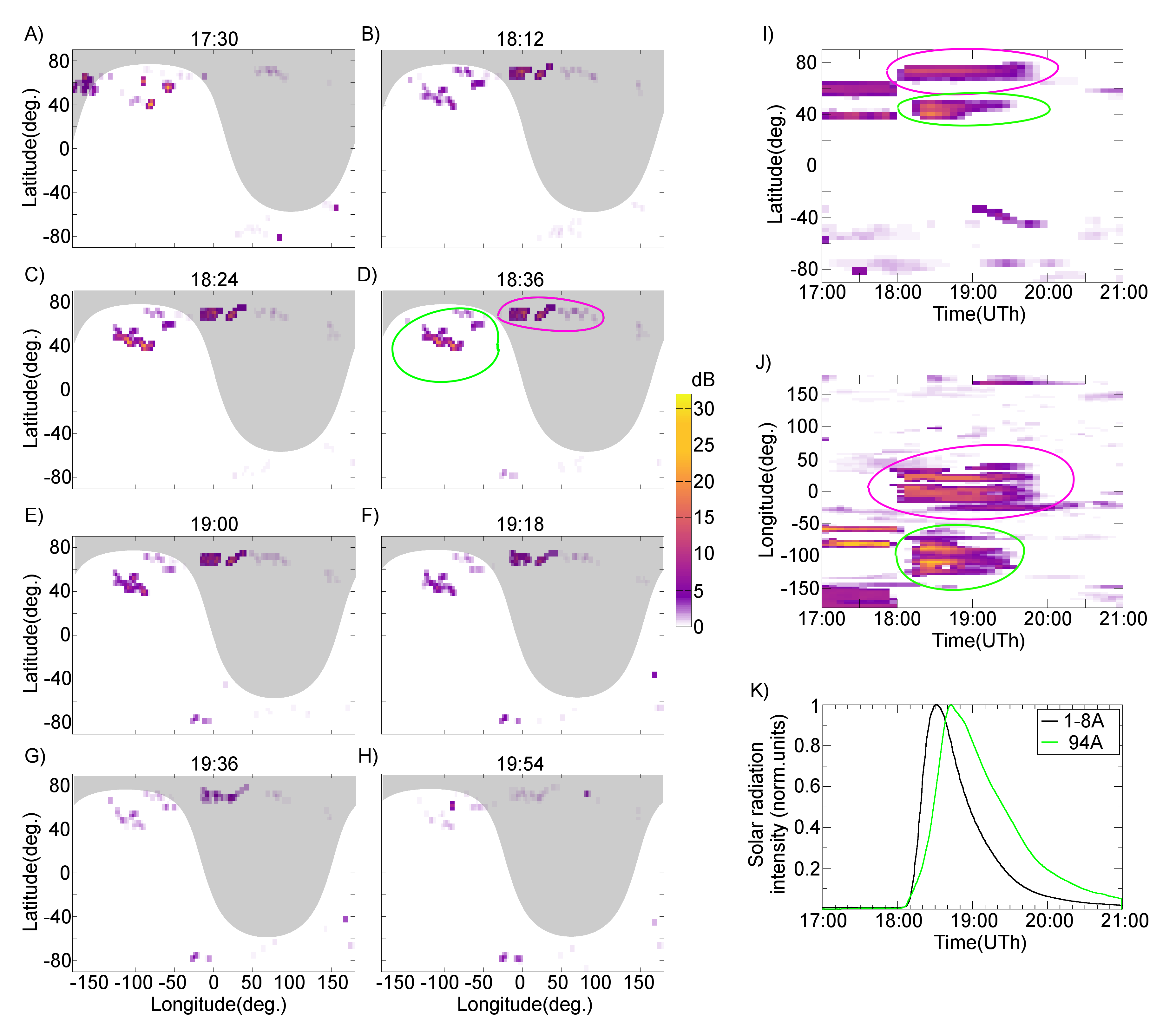}
\caption{A-H) - vertical absorption dynamics at 10MHz during solar X-ray
flare X1.2 07/01/2014 according to the radar network and model (\ref{eq:final_eq}).
Grey region marks unlit area at 100km height.
I) - latitude absorption dynamics during the flare, median over all the longitudes;
J) - longitude absorption dynamics during the flare, median over all the latitudes; 
K) the intensity of solar radiation from the data of GOES/XRS 1-8$\mathring{A}$ and
SDO/AIA 94$\mathring{A}$. Color scale is the same for the figures A-J). Green and violet regions mark effects in  lit and unlit conditions.
}
\label{fig:illustr_goes_wave}
\label{fig:illustr_dynamics}
\end{figure}

Fig.\ref{fig:illustr_dynamics}A-H shows the absorption dynamics over the radars field-of-views
during 
the 07/01/2014 solar flare based on the proposed algorithm. One can see the global-scale
 absorption effect between 18:18~UT and 19:12~UT that 
corresponds to the solar X-ray flare. Each radar produces several measurement points, corresponding to the number of beams, one beam - one measurement point.

So the spatial resolution and resolved areas depend on radiowave propagation characteristics and could vary from flare to flare. 
For future practical purposes one can fit the obtained absorption 
measurements over space by a smoothing function or join them with regular riometric measurements.

One of the ways to smooth the 
obtained data is through their accumulation
 over latitude or longitude. 
It allows us to more clearly distinguish the temporal dynamics of absorption and to reveal its average 
latitudinal 
or longitudinal dependencies.

Fig.\ref{fig:illustr_goes_wave}I shows the dynamics of median absorption
as a function of latitude during this event. The median was calculated over 
3 geographical degrees. 
Fig.\ref{fig:illustr_goes_wave}J shows the dynamics of median absorption
as a function of longitude during this event. The median was calculated over 
3 geographical degrees. For comparison solar radiation at 1-8$\mathring{A}$ and 94$\mathring{A}$ is shown in Fig.\ref{fig:illustr_goes_wave}K.
It can be seen from the figure that the proposed method
makes it possible to investigate the spatio-temporal dynamics of absorption
over a significant part of the Earth's surface. 

A joint analysis of Fig.\ref{fig:illustr_goes_wave}A-J allows, for example,
to distinguish absorption regions in the lit area that correlate well with the flare (green regions) 
from the effects in the unlit area that can not be correctly interpreted 
with the approach taken in this 
paper. The system that we have constructed can be used 
for studies of spatio-temporal
features of daytime absorption both as a separate network and with other instruments and techniques.

\section{Conclusion}

In the present work, a joint analysis was carried out of the data of 
35 HF over-the-horizon radars (34 SuperDARN radars and the EKB ISTP SB RAS radar) during
80 solar flares of 2013-2017. The analysis shows the following features
of the absorption of 8-20MHz radio noise.

The position of an effective noise source on the ground and the error
in determining its location can be determined by the position of spatial
focusing at the boundary of the dead zone and the form of this focusing
(ground scatter signal). This allows using the GS signal to estimate
the position of the region that makes the main contribution to the
observed absorption of the HF radio noise at a particular radar frequency.

The analysis of the correlation between different solar radiation
lines and HF noise dynamics has shown that the temporal variation of the absorption
is well described by a linear combination of the solar radiation intensity
at the wavelengths 1-8$\mathring{A}$ measured by GOES/XRS
and at the wavelength of 94$\mathring{A}$ measured by SDO/AIA.
This allows us to conclude that the main absorption is caused by ionospheric D and E layers.
The assumption we used in our paper about a linear superposition of the contributions 
of each solar 
line to absoprtion is relatively rough.  To solve more accurately for the reconstruction of the electron
 density profile from the experimentally observed noise absorption 
and from the solar spectrum, it is necessary to take into account the processes of ionization by the
various radiation components and corresponding delays more correctly, for example, following the approach of  \cite{Eccles_2005}.

The frequency dependence of the HF absorption is determined by the
median dependence $A[dB]\sim f^{-1.6 \pm 0.3}$.

A model and algorithms are constructed (\ref{eq:final_eq}), that
provides automatic radar estimates of vertical daytime absorption at
10~MHz. Using these model algorithms, it is possible to make statistical
analysis and case-studies of the spatio-temporal dynamics of the absorption of HF
radio 
waves globally, within the coverage area of radar field-of-views.
Each radar produces several measurement points, corresponding to the number of beams, one beam - one measurement point. 
So the spatial resolution and resolved areas depend on radiowave propagation characteristics and will vary from flare to flare. 

One important problem with the algorithm constructed here is the determination of the geographical location of the absorption region 
during the day. This location depends on whether the most intense 1-hop absorption is located near the radar or near the GS distance of 
the first hop.  A similar problem arises with the URSI A1 method.  For future applications, one might want to fit the retrieved absorption measurements
through the use of a smoothing function over space. However, at night or near the terminator, this algorithm should not be used.

Another problem of the algorithm is the impossibility of taking into account irregular 
variations in the background ionosphere.
This is important 
for a more correct estimation of ray trajectory and, as result, for more accurate estimation of the 
vertical absorption from the experimental data for specific observations. The use of calibrated experimental mesurements 
of the  ray elevation angles of GS signals and new techniques for identifying GS signals from radar data should help to solve this problem in the future.

\section*{Acknowledgments}
The data of the SuperDARN radars were obtained using the DaViT (https://github.com /vtsuperdarn /davitpy),
the EKB ISTP SB RAS radar data are the property of the ISTP SB RAS,
contact Oleg Berngardt (berng@iszf.irk.ru). The authors are grateful
to all the developers of the DaViT system, in particular K.Sterne,
N.Frissel, S. de Larquier and A.J.Ribeiro, as well as to all the organizations
supporting the radars operation. O.B. is grateful to X.Shi (Virginia Tech) for 
help in using DaViT.
In the paper we used the data of
EKB ISTP SB RAS, operating under financial support of FSR Program
II.12.2. 
The authors acknowledge the use of SuperDARN data. SuperDARN is a collection of radars funded 
by the national scientific funding agencies of Australia, Canada, China, France, Italy, Japan, 
Norway, South Africa, United Kingdom and United States of America.
The SuperDARN Kerguelen radar is operated by IRAP (CNRS and Toulouse 
University) and is funded by IPEV and CNRS through INSU and PNST programs.
The Dome C East radar was installed in the framework of a French-Italian
collaboration and is operated by INAF-IAPS with the support of CNR and PNRA.
The SuperDARN SENSU Syowa East and South radars are the property of
National Institute of Polar Research (NIPR), this study is a part of the
Science Program of Japanese Antarctic Research Expedition (JARE) and is
supported by NIPR under Ministry of Education, Culture, Sports, Science and
technology (MEXT), Japan.  
The SuperDARN Canada radar operations (SAS, PGR, INV, RKN, CLY) are supported 
by the Canada Foundation for Innovation, the Canadian Space Agency, and the 
Province of Saskatchewan.  The authors thank SuperDARN Canada for
providing the data from the two-frequency operating modes.

The authors are grateful to Altyntsev A.T., Tashchilin A.V., 
Kashapova L.K. (ISTP SB RAS) for useful discussions. The authors are grateful to NOAA
for GOES/XRS and GOES/EUVS data (available at https://satdat.ngdc.noaa.gov /sem /goes /data ), 
to  NASA/SDO and to the AIA and EVE teams for SDO/AIA and SDO/EVE data (available at https://sdo.gsfc.nasa.gov /data/, 
http://lasp.colorado.edu /eve /data\_access /service /file\_download/,
http://suntoday.lmsal.com /suntoday/), to Royal Observatory of Belgium for PROBA2/LYRA data 
(available at http://proba2.oma.be /lyra /data /bsd/) used for analysis. 
The authors are grateful to The University of Southern California Space Sciences Center for using 
SOHO/SEM data, available at https://dornsifecms.usc.edu /space-sciences-center /download-sem-data/). 
Solar Heliospheric Observatory (SOHO) is a joint mission project of United States 
National Aeronautics and Space Administration (NASA) and European Space Agency (ESA).
LYRA is a project of 
the Centre Spatial de Liege, the Physikalisch-Meteorologisches 
Observatorium Davos and the Royal Observatory of Belgium funded by the Belgian Federal Science 
Policy Office (BELSPO) and by the Swiss Bundesamt f\"{u}r Bildung und Wissenschaft.
A.S.Y. is supported by Japan Society for the
Promotion of Science (JSPS), "Grant-in-Aid for Scientific Research (B)"
(Grant Number: 25287129). J.M.R. acknowledges the support of NSF through award AGS-1341918. 
O.I.B. is supported by RFBR grant \#18-05-00539a.




\begin{table}
\caption{The radars participating in the study}
\centering
\begin{tabular}{|c|c|c|p{5cm}|}
\hline 
Code & Geogr.coord. & Full radar name & Owner \\
\hline 
\hline 
ADE & 51.9N,176.6W & Adak Island East SuperDARN & University of Alaska, Fairbanks, USA \\
\hline 
ADW & 51.9N,176.6W & Adak Island West SuperDARN & University of Alaska, Fairbanks, USA \\
\hline 
BKS & 37.1N,77.9W & Blackstone SuperDARN & Virginia Tech, USA \\
\hline 
BPK & 34.6S, 138.5W & Buckland Park SuperDARN & La Trobe University, Australia \\
\hline 
CLY & 70.5N,68.5W & Clyde River SuperDARN & University of Saskatchewan, Canada \\
\hline 
CVE & 43.3N,120.4W & Christmas Valley East SuperDARN & Dartmouth College, USA \\
\hline 
CVW & 43.3N,120.4W & Christmas Valley West SuperDARN & Dartmouth College, USA \\
\hline 
DCE & 75.1S,123.3E & Dome C East SuperDARN & INAF-IAPS/CNR/PNRA, Italy \\
\hline 
EKB & 56.5N,58.5E & Ekaterinburg ISTP SB RAS & ISTP SB RAS, Russia \\
\hline 
FHE & 38.9N, 99.4W & Fort Hays East SuperDARN & Virginia Tech, USA \\
\hline 
FHW & 38.9N, 99.4W & Fort Hays West SuperDARN & Virginia Tech, USA \\
\hline 
GBR & 53.3N,60.5W & Goose Bay SuperDARN & Virginia Tech, USA \\
\hline 
HAL & 75.5S, 75.5W & Halley SuperDARN & British Antarctic Survey, UK \\
\hline 
HAN & 62.3N,26.6E & Hankasalmi SuperDARN & University of Leicester, UK \\
\hline 
HKW & 43.5N,143.6E & Hokkaido West SuperDARN & Nagoya University, Japan \\
\hline 
HOK & 43.5N,143.6E & Hokkaido East SuperDARN & Nagoya University, Japan \\
\hline 
INV & 68.4N,133.8W & Inuvik SuperDARN & University of Saskatchewan, Canada \\
\hline 
KAP & 49.4N,82.3W & Kapuskasing SuperDARN & Virginia Tech, USA \\
\hline 
KER & 49.2S, 70.1E & Kerguelen SuperDARN & IRAP/CNRS/IPEV, France \\
\hline 
KOD & 57.6N,152.2W & Kodiak SuperDARN & University of Alaska, Fairbanks, USA \\
\hline 
KSR & 58.7N,156.6W & King Salmon SuperDARN & National Institute of Information and Communications Technology, Japan \\
\hline 
MCM & 77.9S,166.7E & McMurdo SuperDARN & University of Alaska, Fairbanks, USA \\
\hline 
PGR & 54.0N,122.6W & Prince George SuperDARN & University of Saskatchewan, Canada \\
\hline 
PYK & 63.7N,20.5W & Pykkvibaer SuperDARN & University of Leicester, UK \\
\hline 
RKN & 62.8N,93.1W & Rankin Inlet SuperDARN & University of Saskatchewan, Canada \\
\hline 
SAN & 71.7S, 2.9W & SANAE SuperDARN & South African National Space Agency, South Africa \\
\hline 
SAS & 52.2N,106.5W & Saskatoon SuperDARN & University of Saskatchewan, Canada \\
\hline 
SPS & -90.0S,118.3W & South Pole Station SuperDARN & University of Alaska, Fairbanks, USA \\
\hline 
STO & 63.9N,21.0W & Stokkseyri SuperDARN & Lancaster University, UK \\
\hline 
SYE & 69.0S, 39.6E & Syowa East SuperDARN & National Institute of Polar Research, Japan \\
\hline 
SYS & 69.0S, 39.6E & Syowa South SuperDARN & National Institute of Polar Research, Japan \\
\hline 
TIG & 43.4S, 147.2E & Tiger SuperDARN & La Trobe University, Australia \\
\hline 
UNW & 46.5S, 168.4E & Unwin SuperDARN & La Trobe University, Australia \\
\hline 
WAL & 37.9N,75.5W & Wallops Island SuperDARN & JHU Applied Physics Laboratory, USA \\
\hline 
ZHO & 69.4S,76.4E & Zhongshan SuperDARN & Polar Research Institute of China \\
\hline 
\end{tabular}
\end{table}

\begin{table}
\caption{Spectral bands of solar instruments used in the paper
and their reference average wavelength used in the paper} 

\centering
\begin{tabular}{|c|c|c|}
\hline 
Satellite/Instrument & Spectral band & Reference wavelength ($\mathring{A}$) \\
\hline 
\hline 
GOES/XRS & 1-8$\mathring{A}$ & 5 \\
\hline 
GOES/EUVA & 50-150$\mathring{A}$ & 100 \\
\hline 
GOES/EUVB & 250-400$\mathring{A}$ & 325 \\
\hline 
GOES/EUVC & 200-700$\mathring{A}$ & 450 \\
\hline 
GOES/EUVD & 200-850$\mathring{A}$ & 525 \\
\hline 
GOES/EUVE & 1150-1250$\mathring{A}$ & 1200 \\
\hline 
SDO/AIA & 94$\mathring{A}$ & 94 \\
\hline 
SDO/AIA & 131$\mathring{A}$ & 131 \\
\hline 
SDO/AIA & 171$\mathring{A}$ & 171 \\
\hline 
SDO/AIA & 193$\mathring{A}$ & 193 \\
\hline 
SDO/AIA & 211$\mathring{A}$ & 211 \\
\hline 
SDO/AIA & 304$\mathring{A}$ & 304 \\
\hline 
SDO/AIA & 335$\mathring{A}$ & 335 \\
\hline 
SDO/AIA & 1600$\mathring{A}$ & 1600 \\
\hline 
SDO/AIA & 1700$\mathring{A}$ & 1700 \\
\hline 
PROBA2/LYRA (channel 1) & 1200-1230$\mathring{A}$ & 1215 \\
\hline 
PROBA2/LYRA (channel 2) & 1900-2220$\mathring{A}$ & 2060 \\
\hline 
PROBA2/LYRA (channel 3) &  $<$50$\mathring{A}$ + 170-800$\mathring{A}$ & 435 \\
\hline 
PROBA2/LYRA (channel 4) &  $<$20$\mathring{A}$ + 60-200$\mathring{A}$ & 130 \\
\hline
SOHO/SEM (channel 2) & 1-500$\mathring{A}$ & 249 \\
\hline 
SOHO/SEM (channels 1+3) & 260-340$\mathring{A}$ & 300 \\
\hline 
SDO/EVE (ESP) &  1-70$\mathring{A}$ & 35 \\
\hline 
\end{tabular}
\label{tab:sol_wl}
\end{table}

\bibliographystyle{apacite}

\begin{thebibliography}{}

\bibitem [\protect \citeauthoryear {%
{Akmaev, R. A.}%
}{%
{Akmaev, R. A.}%
}{%
{\protect \APACyear {2010}}%
}]{%
DRAP2}
\APACinsertmetastar {%
DRAP2}%
\begin{APACrefauthors}%
{Akmaev, R. A.}%
\end{APACrefauthors}%
\unskip\
\newblock
\APACrefYearMonthDay{2010}{}{}.
\newblock
\APACrefbtitle {{DRAP Model Validation: I.Scientific Report}.} {{DRAP Model
  Validation: I.Scientific Report}.}
\newblock
\begin{APACrefURL} \url{https://www.ngdc.noaa.gov/stp/drap/DRAP-V-Report1.pdf}
  \end{APACrefURL}
\PrintBackRefs{\CurrentBib}

\bibitem [\protect \citeauthoryear {%
Bailey%
}{%
Bailey%
}{%
{\protect \APACyear {1994}}%
}]{%
bailey1994}
\APACinsertmetastar {%
bailey1994}%
\begin{APACrefauthors}%
Bailey, K.%
\end{APACrefauthors}%
\unskip\
\newblock
\APACrefYear{1994}.
\newblock
\APACrefbtitle {{Typologies and Taxonomies: An Introduction to Classification
  Techniques}} {{Typologies and Taxonomies: An Introduction to Classification
  Techniques}}\ (\BNUM~102).
\newblock
\APACaddressPublisher{}{SAGE Publications}.
\PrintBackRefs{\CurrentBib}

\bibitem [\protect \citeauthoryear {%
J.~Baker%
\ \protect \BOthers {.}}{%
J.~Baker%
\ \protect \BOthers {.}}{%
{\protect \APACyear {2007}}%
}]{%
Baker_2007}
\APACinsertmetastar {%
Baker_2007}%
\begin{APACrefauthors}%
Baker, J.%
, Greenwald, R.%
, Ruohoniemi, J.%
, Oksavik, K.%
, Gjerloev, J\BPBI W.%
, Paxton, L\BPBI J.%
\BCBL {}\ \BBA {} Hairston, M.%
\end{APACrefauthors}%
\unskip\
\newblock
\APACrefYearMonthDay{2007}{}{}.
\newblock
{\BBOQ}\APACrefatitle {{Observations of ionospheric convection from the Wallops
  SuperDARN radar at middle latitudes}} {{Observations of ionospheric
  convection from the Wallops SuperDARN radar at middle latitudes}}.{\BBCQ}
\newblock
\APACjournalVolNumPages{Journal of Geophysical Research (Space
  Physics)}{112}{}{A01303}.
\newblock
\begin{APACrefDOI} \doi{10.1029/2006ja011982} \end{APACrefDOI}
\PrintBackRefs{\CurrentBib}

\bibitem [\protect \citeauthoryear {%
K\BPBI B.~Baker%
, Greenwald%
, Villian%
\BCBL {}\ \BBA {} Wing%
}{%
K\BPBI B.~Baker%
\ \protect \BOthers {.}}{%
{\protect \APACyear {1988}}%
}]{%
Baker_1988}
\APACinsertmetastar {%
Baker_1988}%
\begin{APACrefauthors}%
Baker, K\BPBI B.%
, Greenwald, R.%
, Villian, J\BPBI P.%
\BCBL {}\ \BBA {} Wing, S.%
\end{APACrefauthors}%
\unskip\
\newblock
\APACrefYearMonthDay{1988}{}{}.
\newblock
\APACrefbtitle {{Spectral Characteristics of High Frequency (HF) Backscatter
  for High Latitude Ionospheric Irregularities: Preliminary Analysis of
  Statistical Properties}} {{Spectral Characteristics of High Frequency (HF)
  Backscatter for High Latitude Ionospheric Irregularities: Preliminary
  Analysis of Statistical Properties}}\ \APACbVolEdTR{}{\BTR{}\ \BNUM\
  ADA202998}.
\newblock
\APACaddressInstitution{}{Johns Hopkins Univ Laurel Md Applied Physics Lab}.
\PrintBackRefs{\CurrentBib}

\bibitem [\protect \citeauthoryear {%
Banks%
\ \BBA {} Kockarts%
}{%
Banks%
\ \BBA {} Kockarts%
}{%
{\protect \APACyear {1973}}%
}]{%
banks1973aeronomy}
\APACinsertmetastar {%
banks1973aeronomy}%
\begin{APACrefauthors}%
Banks, P.%
\BCBT {}\ \BBA {} Kockarts, G.%
\end{APACrefauthors}%
\unskip\
\newblock
\APACrefYear{1973}.
\newblock
\APACrefbtitle {Aeronomy} {Aeronomy}\ (\BVOL~A).
\newblock
\APACaddressPublisher{}{Academic Press, New York and London}.
\PrintBackRefs{\CurrentBib}

\bibitem [\protect \citeauthoryear {%
Barthes%
, Andre%
, Cerisier%
\BCBL {}\ \BBA {} Villain%
}{%
Barthes%
\ \protect \BOthers {.}}{%
{\protect \APACyear {1998}}%
}]{%
Barthes_1998}
\APACinsertmetastar {%
Barthes_1998}%
\begin{APACrefauthors}%
Barthes, L.%
, Andre, D.%
, Cerisier, J\BPBI C.%
\BCBL {}\ \BBA {} Villain, J\BHBI P.%
\end{APACrefauthors}%
\unskip\
\newblock
\APACrefYearMonthDay{1998}{}{}.
\newblock
{\BBOQ}\APACrefatitle {{Separation of multiple echoes using a high--resolution
  spectral analysis for SuperDARN HF radars}} {{Separation of multiple echoes
  using a high--resolution spectral analysis for SuperDARN HF radars}}.{\BBCQ}
\newblock
\APACjournalVolNumPages{Radio Science}{33}{4}{1005--1017}.
\newblock
\begin{APACrefDOI} \doi{10.1029/98rs00714} \end{APACrefDOI}
\PrintBackRefs{\CurrentBib}

\bibitem [\protect \citeauthoryear {%
Berngardt%
\ \protect \BOthers {.}}{%
Berngardt%
\ \protect \BOthers {.}}{%
{\protect \APACyear {2018}}%
}]{%
BERNGARDT_20181}
\APACinsertmetastar {%
BERNGARDT_20181}%
\begin{APACrefauthors}%
Berngardt, O\BPBI I.%
, Ruohoniemi, J\BPBI M.%
, Nishitani, N.%
, Shepherd, S\BPBI G.%
, Bristow, W\BPBI A.%
\BCBL {}\ \BBA {} Miller, E\BPBI S.%
\end{APACrefauthors}%
\unskip\
\newblock
\APACrefYearMonthDay{2018}{}{}.
\newblock
{\BBOQ}\APACrefatitle {{Attenuation of decameter wavelength sky noise during
  x--ray solar flares in 2013--2017 based on the observations of midlatitude HF
  radars}} {{Attenuation of decameter wavelength sky noise during x--ray solar
  flares in 2013--2017 based on the observations of midlatitude HF
  radars}}.{\BBCQ}
\newblock
\APACjournalVolNumPages{Journal of Atmospheric and Solar-Terrestrial
  Physics}{173}{}{1 - 13}.
\newblock
\begin{APACrefDOI} \doi{10.1016/j.jastp.2018.03.022} \end{APACrefDOI}
\PrintBackRefs{\CurrentBib}

\bibitem [\protect \citeauthoryear {%
Berngardt%
, Zolotukhina%
\BCBL {}\ \BBA {} Oinats%
}{%
Berngardt%
\ \protect \BOthers {.}}{%
{\protect \APACyear {2015}}%
}]{%
Berngardt_2015}
\APACinsertmetastar {%
Berngardt_2015}%
\begin{APACrefauthors}%
Berngardt, O\BPBI I.%
, Zolotukhina, N\BPBI A.%
\BCBL {}\ \BBA {} Oinats, A\BPBI V.%
\end{APACrefauthors}%
\unskip\
\newblock
\APACrefYearMonthDay{2015}{}{}.
\newblock
{\BBOQ}\APACrefatitle {{Observations of field--aligned ionospheric
  irregularities during quiet and disturbed conditions with EKB radar: first
  results}} {{Observations of field--aligned ionospheric irregularities during
  quiet and disturbed conditions with EKB radar: first results}}.{\BBCQ}
\newblock
\APACjournalVolNumPages{Earth, Planets and Space}{67}{1}{143}.
\newblock
\begin{APACrefDOI} \doi{10.1186/s40623-015-0302-3} \end{APACrefDOI}
\PrintBackRefs{\CurrentBib}

\bibitem [\protect \citeauthoryear {%
Bilitza%
\ \protect \BOthers {.}}{%
Bilitza%
\ \protect \BOthers {.}}{%
{\protect \APACyear {2017}}%
}]{%
Bilitza_2017}
\APACinsertmetastar {%
Bilitza_2017}%
\begin{APACrefauthors}%
Bilitza, D.%
, Altadill, D.%
, Truhlik, V.%
, Shubin, V.%
, Galkin, I\BPBI A.%
, Reinisch, B\BPBI W.%
\BCBL {}\ \BBA {} Huang, X.%
\end{APACrefauthors}%
\unskip\
\newblock
\APACrefYearMonthDay{2017}{}{}.
\newblock
{\BBOQ}\APACrefatitle {{International Reference Ionosphere 2016: from
  ionospheric climate to real--time weather predictions}} {{International
  Reference Ionosphere 2016: from ionospheric climate to real--time weather
  predictions}}.{\BBCQ}
\newblock
\APACjournalVolNumPages{Space Weather}{}{}{418--429}.
\newblock
\APACrefnote{2016SW001593}
\newblock
\begin{APACrefDOI} \doi{10.1002/2016sw001593} \end{APACrefDOI}
\PrintBackRefs{\CurrentBib}

\bibitem [\protect \citeauthoryear {%
Blagoveshchenskii%
, Maltseva%
, Anishin%
, Rogov%
\BCBL {}\ \BBA {} Sergeeva%
}{%
Blagoveshchenskii%
\ \protect \BOthers {.}}{%
{\protect \APACyear {2015}}%
}]{%
Blagov_2015}
\APACinsertmetastar {%
Blagov_2015}%
\begin{APACrefauthors}%
Blagoveshchenskii, D\BPBI V.%
, Maltseva, O\BPBI A.%
, Anishin, M\BPBI M.%
, Rogov, D\BPBI D.%
\BCBL {}\ \BBA {} Sergeeva, M\BPBI A.%
\end{APACrefauthors}%
\unskip\
\newblock
\APACrefYearMonthDay{2015}{}{}.
\newblock
{\BBOQ}\APACrefatitle {{Modeling of HF propagation at high latitudes on the
  basis of IRI}} {{Modeling of HF propagation at high latitudes on the basis of
  IRI}}.{\BBCQ}
\newblock
\APACjournalVolNumPages{Advances in Space Research}{57}{3}{821-834}.
\newblock
\begin{APACrefDOI} \doi{10.1016/j.asr.2015.11.029} \end{APACrefDOI}
\PrintBackRefs{\CurrentBib}

\bibitem [\protect \citeauthoryear {%
Blanchard%
, Sundeen%
\BCBL {}\ \BBA {} Baker%
}{%
Blanchard%
\ \protect \BOthers {.}}{%
{\protect \APACyear {2009}}%
}]{%
Blanchard_2009}
\APACinsertmetastar {%
Blanchard_2009}%
\begin{APACrefauthors}%
Blanchard, G\BPBI T.%
, Sundeen, S.%
\BCBL {}\ \BBA {} Baker, K\BPBI B.%
\end{APACrefauthors}%
\unskip\
\newblock
\APACrefYearMonthDay{2009}{}{}.
\newblock
{\BBOQ}\APACrefatitle {Probabilistic identification of high--frequency radar
  backscatter from the ground and ionosphere based on spectral characteristics}
  {Probabilistic identification of high--frequency radar backscatter from the
  ground and ionosphere based on spectral characteristics}.{\BBCQ}
\newblock
\APACjournalVolNumPages{Radio Science}{44}{5}{RS5012}.
\newblock
\begin{APACrefDOI} \doi{10.1029/2009rs004141} \end{APACrefDOI}
\PrintBackRefs{\CurrentBib}

\bibitem [\protect \citeauthoryear {%
Bland%
, Heino%
, Kosch%
\BCBL {}\ \BBA {} Partamies%
}{%
Bland%
\ \protect \BOthers {.}}{%
{\protect \APACyear {2018}}%
}]{%
Bland_2018}
\APACinsertmetastar {%
Bland_2018}%
\begin{APACrefauthors}%
Bland, E\BPBI C.%
, Heino, E.%
, Kosch, M\BPBI J.%
\BCBL {}\ \BBA {} Partamies, N.%
\end{APACrefauthors}%
\unskip\
\newblock
\APACrefYearMonthDay{2018}{}{}.
\newblock
{\BBOQ}\APACrefatitle {{SuperDARN radar--derived HF radio attenuation during
  the September 2017 solar proton events}} {{SuperDARN radar--derived HF radio
  attenuation during the September 2017 solar proton events}}.{\BBCQ}
\newblock
\APACjournalVolNumPages{Space Weather}{}{}{}.
\newblock
\begin{APACrefDOI} \doi{10.1029/2018sw001916} \end{APACrefDOI}
\PrintBackRefs{\CurrentBib}

\bibitem [\protect \citeauthoryear {%
Bliokh%
, Galushko%
, Minakov%
\BCBL {}\ \BBA {} Yampolski%
}{%
Bliokh%
\ \protect \BOthers {.}}{%
{\protect \APACyear {1988}}%
}]{%
Bliokh1988}
\APACinsertmetastar {%
Bliokh1988}%
\begin{APACrefauthors}%
Bliokh, P\BPBI V.%
, Galushko, V\BPBI G.%
, Minakov, A\BPBI A.%
\BCBL {}\ \BBA {} Yampolski, Y\BPBI M.%
\end{APACrefauthors}%
\unskip\
\newblock
\APACrefYearMonthDay{1988}{}{}.
\newblock
{\BBOQ}\APACrefatitle {Field interference structure fluctuations near the
  boundary of the skip zone} {Field interference structure fluctuations near
  the boundary of the skip zone}.{\BBCQ}
\newblock
\APACjournalVolNumPages{Radiophysics and Quantum Electronics}{31}{6}{480--487}.
\newblock
\begin{APACrefDOI} \doi{10.1007/bf01044650} \end{APACrefDOI}
\PrintBackRefs{\CurrentBib}

\bibitem [\protect \citeauthoryear {%
Chakraborty%
, Ruohoniemi%
, Baker%
\BCBL {}\ \BBA {} Nishitani%
}{%
Chakraborty%
\ \protect \BOthers {.}}{%
{\protect \APACyear {2018}}%
}]{%
Chakraborty_2018}
\APACinsertmetastar {%
Chakraborty_2018}%
\begin{APACrefauthors}%
Chakraborty, S.%
, Ruohoniemi, J\BPBI M.%
, Baker, J\BPBI B\BPBI H.%
\BCBL {}\ \BBA {} Nishitani, N.%
\end{APACrefauthors}%
\unskip\
\newblock
\APACrefYearMonthDay{2018}{}{}.
\newblock
{\BBOQ}\APACrefatitle {{Characterization of Short--Wave Fadeout Seen in Daytime
  SuperDARN Ground Scatter Observations}} {{Characterization of Short--Wave
  Fadeout Seen in Daytime SuperDARN Ground Scatter Observations}}.{\BBCQ}
\newblock
\APACjournalVolNumPages{Radio Science}{53}{4}{472-484}.
\newblock
\begin{APACrefDOI} \doi{10.1002/2017RS006488} \end{APACrefDOI}
\PrintBackRefs{\CurrentBib}

\bibitem [\protect \citeauthoryear {%
Chernov%
}{%
Chernov%
}{%
{\protect \APACyear {1971}}%
}]{%
Chernov1971}
\APACinsertmetastar {%
Chernov1971}%
\begin{APACrefauthors}%
Chernov, Y\BPBI A.%
\end{APACrefauthors}%
\unskip\
\newblock
\APACrefYear{1971}.
\newblock
\APACrefbtitle {Backward--oblique sounding of the ionosphere(in russian)}
  {Backward--oblique sounding of the ionosphere(in russian)}.
\newblock
\APACaddressPublisher{}{Moscow,Svyaz}.
\PrintBackRefs{\CurrentBib}

\bibitem [\protect \citeauthoryear {%
Chisham%
}{%
Chisham%
}{%
{\protect \APACyear {2018}}%
}]{%
Chisham_2018}
\APACinsertmetastar {%
Chisham_2018}%
\begin{APACrefauthors}%
Chisham, G.%
\end{APACrefauthors}%
\unskip\
\newblock
\APACrefYearMonthDay{2018}{}{}.
\newblock
{\BBOQ}\APACrefatitle {{Calibrating SuperDARN Interferometers Using Meteor
  Backscatter}} {{Calibrating SuperDARN Interferometers Using Meteor
  Backscatter}}.{\BBCQ}
\newblock
\APACjournalVolNumPages{Radio Science}{53}{6}{761-774}.
\newblock
\begin{APACrefDOI} \doi{10.1029/2017RS006492} \end{APACrefDOI}
\PrintBackRefs{\CurrentBib}

\bibitem [\protect \citeauthoryear {%
Chisham%
\ \protect \BOthers {.}}{%
Chisham%
\ \protect \BOthers {.}}{%
{\protect \APACyear {2007}}%
}]{%
Chisham_2007}
\APACinsertmetastar {%
Chisham_2007}%
\begin{APACrefauthors}%
Chisham, G.%
, Lester, M.%
, Milan, S.%
, Freeman, M.%
, Bristow, W.%
, McWilliams, K.%
\BDBL {}Walker, A.%
\end{APACrefauthors}%
\unskip\
\newblock
\APACrefYearMonthDay{2007}{}{}.
\newblock
{\BBOQ}\APACrefatitle {{A decade of the Super Dual Auroral Radar Network
  (SuperDARN): scientific achievements, new techniques and future directions}}
  {{A decade of the Super Dual Auroral Radar Network (SuperDARN): scientific
  achievements, new techniques and future directions}}.{\BBCQ}
\newblock
\APACjournalVolNumPages{Surveys in Geophysics}{}{28}{33-109}.
\newblock
\begin{APACrefDOI} \doi{10.1007/s10712-007-9017-8} \end{APACrefDOI}
\PrintBackRefs{\CurrentBib}

\bibitem [\protect \citeauthoryear {%
Davies%
}{%
Davies%
}{%
{\protect \APACyear {1969}}%
}]{%
Davies_1969}
\APACinsertmetastar {%
Davies_1969}%
\begin{APACrefauthors}%
Davies, K.%
\end{APACrefauthors}%
\unskip\
\newblock
\APACrefYear{1969}.
\newblock
\APACrefbtitle {Ionospheric radio waves} {Ionospheric radio waves}.
\newblock
\APACaddressPublisher{}{Blaisdell Pub. Co.}
\PrintBackRefs{\CurrentBib}

\bibitem [\protect \citeauthoryear {%
Detrick%
\ \BBA {} Rosenberg%
}{%
Detrick%
\ \BBA {} Rosenberg%
}{%
{\protect \APACyear {1990}}%
}]{%
Detrick_1990}
\APACinsertmetastar {%
Detrick_1990}%
\begin{APACrefauthors}%
Detrick, D\BPBI L.%
\BCBT {}\ \BBA {} Rosenberg, T\BPBI J.%
\end{APACrefauthors}%
\unskip\
\newblock
\APACrefYearMonthDay{1990}{}{}.
\newblock
{\BBOQ}\APACrefatitle {A phased--array radiowave imager for studies of cosmic
  noise absorption} {A phased--array radiowave imager for studies of cosmic
  noise absorption}.{\BBCQ}
\newblock
\APACjournalVolNumPages{Radio Science}{25}{4}{325-338}.
\newblock
\begin{APACrefDOI} \doi{10.1029/RS025i004p00325} \end{APACrefDOI}
\PrintBackRefs{\CurrentBib}

\bibitem [\protect \citeauthoryear {%
Didkovsky%
, Judge%
, Wieman%
, Woods%
\BCBL {}\ \BBA {} Jones%
}{%
Didkovsky%
\ \protect \BOthers {.}}{%
{\protect \APACyear {2012}}%
}]{%
Didkovsky_2012}
\APACinsertmetastar {%
Didkovsky_2012}%
\begin{APACrefauthors}%
Didkovsky, L\BPBI V.%
, Judge, D.%
, Wieman, S.%
, Woods, T.%
\BCBL {}\ \BBA {} Jones, A.%
\end{APACrefauthors}%
\unskip\
\newblock
\APACrefYearMonthDay{2012}{}{}.
\newblock
{\BBOQ}\APACrefatitle {{EUV SpectroPhotometer (ESP) in Extreme Ultraviolet
  Variability Experiment (EVE): Algorithms and Calibrations}} {{EUV
  SpectroPhotometer (ESP) in Extreme Ultraviolet Variability Experiment (EVE):
  Algorithms and Calibrations}}.{\BBCQ}
\newblock
\APACjournalVolNumPages{Solar Physics}{275}{1}{179--205}.
\newblock
\begin{APACrefDOI} \doi{10.1007/s11207-009-9485-8} \end{APACrefDOI}
\PrintBackRefs{\CurrentBib}

\bibitem [\protect \citeauthoryear {%
Didkovsky%
\ \protect \BOthers {.}}{%
Didkovsky%
\ \protect \BOthers {.}}{%
{\protect \APACyear {2006}}%
}]{%
Didkovsky_2006}
\APACinsertmetastar {%
Didkovsky_2006}%
\begin{APACrefauthors}%
Didkovsky, L\BPBI V.%
, Judge, D\BPBI L.%
, Jones, A\BPBI R.%
, Wieman, S.%
, Tsurutani, B\BPBI T.%
\BCBL {}\ \BBA {} McMullin, D.%
\end{APACrefauthors}%
\unskip\
\newblock
\APACrefYearMonthDay{2006}{}{}.
\newblock
{\BBOQ}\APACrefatitle {{Correction of SOHO CELIAS/SEM EUV measurements
  saturated by extreme solar flare events}} {{Correction of SOHO CELIAS/SEM EUV
  measurements saturated by extreme solar flare events}}.{\BBCQ}
\newblock
\APACjournalVolNumPages{Astronomische Nachrichten}{328}{1}{36-40}.
\newblock
\begin{APACrefDOI} \doi{10.1002/asna.200610667} \end{APACrefDOI}
\PrintBackRefs{\CurrentBib}

\bibitem [\protect \citeauthoryear {%
Dominique%
\ \protect \BOthers {.}}{%
Dominique%
\ \protect \BOthers {.}}{%
{\protect \APACyear {2013}}%
}]{%
Dominique_2013}
\APACinsertmetastar {%
Dominique_2013}%
\begin{APACrefauthors}%
Dominique, M.%
, Hochedez, J\BHBI F.%
, Schmutz, W.%
, Dammasch, I\BPBI E.%
, Shapiro, A\BPBI I.%
, Kretzschmar, M.%
\BDBL {}BenMoussa, A.%
\end{APACrefauthors}%
\unskip\
\newblock
\APACrefYearMonthDay{2013}{}{}.
\newblock
{\BBOQ}\APACrefatitle {{The LYRA Instrument Onboard PROBA2: Description and
  In--Flight Performance}} {{The LYRA Instrument Onboard PROBA2: Description
  and In--Flight Performance}}.{\BBCQ}
\newblock
\APACjournalVolNumPages{{Solar Physics}}{286}{1}{21--42}.
\newblock
\begin{APACrefDOI} \doi{10.1007/s11207-013-0252-5} \end{APACrefDOI}
\PrintBackRefs{\CurrentBib}

\bibitem [\protect \citeauthoryear {%
Donnelly%
}{%
Donnelly%
}{%
{\protect \APACyear {1976}}%
}]{%
Donnelly_1976}
\APACinsertmetastar {%
Donnelly_1976}%
\begin{APACrefauthors}%
Donnelly, R\BPBI F.%
\end{APACrefauthors}%
\unskip\
\newblock
\APACrefYearMonthDay{1976}{}{}.
\newblock
{\BBOQ}\APACrefatitle {{Empirical models of solar flare X ray and EUV emission
  for use in studying their E and F region effects}} {{Empirical models of
  solar flare X ray and EUV emission for use in studying their E and F region
  effects}}.{\BBCQ}
\newblock
\APACjournalVolNumPages{Journal of Geophysical Research}{81}{25}{4745-4753}.
\newblock
\begin{APACrefDOI} \doi{10.1029/JA081i025p04745} \end{APACrefDOI}
\PrintBackRefs{\CurrentBib}

\bibitem [\protect \citeauthoryear {%
{DRAP~Documentation}%
}{%
{DRAP~Documentation}%
}{%
{\protect \APACyear {2010}}%
}]{%
DRAP}
\APACinsertmetastar {%
DRAP}%
\begin{APACrefauthors}%
{DRAP~Documentation}.%
\end{APACrefauthors}%
\unskip\
\newblock
\APACrefYearMonthDay{2010}{}{}.
\newblock
\APACrefbtitle {{Global D--region absorption prediction documentation, accessed
  September,2018}.} {{Global D--region absorption prediction documentation,
  accessed September,2018}.}
\newblock
\begin{APACrefURL}
  \url{https://www.swpc.noaa.gov/content/global-d-region-absorption-prediction-documentation}
  \end{APACrefURL}
\PrintBackRefs{\CurrentBib}

\bibitem [\protect \citeauthoryear {%
Eccles%
, Hunsucker%
, Rice%
\BCBL {}\ \BBA {} Sojka%
}{%
Eccles%
\ \protect \BOthers {.}}{%
{\protect \APACyear {2005}}%
}]{%
Eccles_2005}
\APACinsertmetastar {%
Eccles_2005}%
\begin{APACrefauthors}%
Eccles, J\BPBI V.%
, Hunsucker, R\BPBI D.%
, Rice, D.%
\BCBL {}\ \BBA {} Sojka, J\BPBI J.%
\end{APACrefauthors}%
\unskip\
\newblock
\APACrefYearMonthDay{2005}{}{}.
\newblock
{\BBOQ}\APACrefatitle {{Space weather effects on midlatitude HF propagation
  paths: Observations and a data--driven D region model}} {{Space weather
  effects on midlatitude HF propagation paths: Observations and a data--driven
  D region model}}.{\BBCQ}
\newblock
\APACjournalVolNumPages{Space Weather}{3}{1}{}.
\newblock
\APACrefnote{S01002}
\newblock
\begin{APACrefDOI} \doi{10.1029/2004sw000094} \end{APACrefDOI}
\PrintBackRefs{\CurrentBib}

\bibitem [\protect \citeauthoryear {%
Fiori%
\ \protect \BOthers {.}}{%
Fiori%
\ \protect \BOthers {.}}{%
{\protect \APACyear {2018}}%
}]{%
Fiori_2018}
\APACinsertmetastar {%
Fiori_2018}%
\begin{APACrefauthors}%
Fiori, R\BPBI A\BPBI D.%
, Koustov, A\BPBI V.%
, Chakraborty, S.%
, Ruohoniemi, J\BPBI M.%
, Danskin, D\BPBI W.%
, Boteler, D\BPBI H.%
\BCBL {}\ \BBA {} Shepherd, S\BPBI G.%
\end{APACrefauthors}%
\unskip\
\newblock
\APACrefYearMonthDay{2018}{}{}.
\newblock
{\BBOQ}\APACrefatitle {{Examining the potential of the Super Dual Auroral Radar
  Network for monitoring the space weather impact of solar X--ray flares}}
  {{Examining the potential of the Super Dual Auroral Radar Network for
  monitoring the space weather impact of solar X--ray flares}}.{\BBCQ}
\newblock
\APACjournalVolNumPages{Space Weather}{}{}{}.
\newblock
\begin{APACrefDOI} \doi{10.1029/2018sw001905} \end{APACrefDOI}
\PrintBackRefs{\CurrentBib}

\bibitem [\protect \citeauthoryear {%
Galkin%
, Reinisch%
, Huang%
\BCBL {}\ \BBA {} Bilitza%
}{%
Galkin%
\ \protect \BOthers {.}}{%
{\protect \APACyear {2012}}%
}]{%
Galkin_2011}
\APACinsertmetastar {%
Galkin_2011}%
\begin{APACrefauthors}%
Galkin, I\BPBI A.%
, Reinisch, B\BPBI W.%
, Huang, X.%
\BCBL {}\ \BBA {} Bilitza, D.%
\end{APACrefauthors}%
\unskip\
\newblock
\APACrefYearMonthDay{2012}{}{}.
\newblock
{\BBOQ}\APACrefatitle {{Assimilation of GIRO data into a real--time IRI}}
  {{Assimilation of GIRO data into a real--time IRI}}.{\BBCQ}
\newblock
\APACjournalVolNumPages{Radio Science}{47}{4}{}.
\newblock
\begin{APACrefDOI} \doi{10.1029/2011RS004952} \end{APACrefDOI}
\PrintBackRefs{\CurrentBib}

\bibitem [\protect \citeauthoryear {%
Greenwald%
\ \protect \BOthers {.}}{%
Greenwald%
\ \protect \BOthers {.}}{%
{\protect \APACyear {1995}}%
}]{%
Greenwald_1995}
\APACinsertmetastar {%
Greenwald_1995}%
\begin{APACrefauthors}%
Greenwald, R.%
, Baker, K\BPBI B.%
, Dudeney, J\BPBI R.%
, Pinnock, M.%
, Jones, T.%
, Thomas, E.%
\BDBL {}Yamagishi, H.%
\end{APACrefauthors}%
\unskip\
\newblock
\APACrefYearMonthDay{1995}{}{}.
\newblock
{\BBOQ}\APACrefatitle {{Darn/Superdarn: A Global View of the Dynamics of
  High--Lattitude Convection}} {{Darn/Superdarn: A Global View of the Dynamics
  of High--Lattitude Convection}}.{\BBCQ}
\newblock
\APACjournalVolNumPages{Space Science Reviews}{71}{}{761--796}.
\newblock
\begin{APACrefDOI} \doi{10.1007/BF00751350} \end{APACrefDOI}
\PrintBackRefs{\CurrentBib}

\bibitem [\protect \citeauthoryear {%
Hall%
\ \protect \BOthers {.}}{%
Hall%
\ \protect \BOthers {.}}{%
{\protect \APACyear {1997}}%
}]{%
Hall_1997}
\APACinsertmetastar {%
Hall_1997}%
\begin{APACrefauthors}%
Hall, G\BPBI E.%
, MacDougall, J\BPBI W.%
, Moorcroft, D\BPBI R.%
, St.-Maurice, J\BHBI P.%
, Manson, A\BPBI H.%
\BCBL {}\ \BBA {} Meek, C\BPBI E.%
\end{APACrefauthors}%
\unskip\
\newblock
\APACrefYearMonthDay{1997}{}{}.
\newblock
{\BBOQ}\APACrefatitle {{Super Dual Auroral Radar Network observations of meteor
  echoes}} {{Super Dual Auroral Radar Network observations of meteor
  echoes}}.{\BBCQ}
\newblock
\APACjournalVolNumPages{Journal of Geophysical Research: Space
  Physics}{102}{A7}{14603-14614}.
\newblock
\begin{APACrefDOI} \doi{10.1029/97JA00517} \end{APACrefDOI}
\PrintBackRefs{\CurrentBib}

\bibitem [\protect \citeauthoryear {%
Hanser%
\ \BBA {} Sellers%
}{%
Hanser%
\ \BBA {} Sellers%
}{%
{\protect \APACyear {1996}}%
}]{%
Hanser_1996}
\APACinsertmetastar {%
Hanser_1996}%
\begin{APACrefauthors}%
Hanser, F\BPBI A.%
\BCBT {}\ \BBA {} Sellers, F\BPBI B.%
\end{APACrefauthors}%
\unskip\
\newblock
\APACrefYearMonthDay{1996}{}{}.
\newblock
{\BBOQ}\APACrefatitle {{Design and calibration of the GOES--8 solar x--ray
  sensor: the XRS}} {{Design and calibration of the GOES--8 solar x--ray
  sensor: the XRS}}.{\BBCQ}
\newblock
\APACjournalVolNumPages{Proc.SPIE}{2812}{}{2812 - 2812-9}.
\newblock
\begin{APACrefDOI} \doi{10.1117/12.254082} \end{APACrefDOI}
\PrintBackRefs{\CurrentBib}

\bibitem [\protect \citeauthoryear {%
Hargreaves%
}{%
Hargreaves%
}{%
{\protect \APACyear {2010}}%
}]{%
Hargreaves_2010}
\APACinsertmetastar {%
Hargreaves_2010}%
\begin{APACrefauthors}%
Hargreaves, J.%
\end{APACrefauthors}%
\unskip\
\newblock
\APACrefYearMonthDay{2010}{}{}.
\newblock
{\BBOQ}\APACrefatitle {Auroral radio absorption: The prediction question}
  {Auroral radio absorption: The prediction question}.{\BBCQ}
\newblock
\APACjournalVolNumPages{Advances in Space Research}{45}{9}{1075 - 1092}.
\newblock
\begin{APACrefDOI} \doi{10.1016/j.asr.2009.10.026} \end{APACrefDOI}
\PrintBackRefs{\CurrentBib}

\bibitem [\protect \citeauthoryear {%
{Hochedez}%
\ \protect \BOthers {.}}{%
{Hochedez}%
\ \protect \BOthers {.}}{%
{\protect \APACyear {2006}}%
}]{%
Hochedez_2006}
\APACinsertmetastar {%
Hochedez_2006}%
\begin{APACrefauthors}%
{Hochedez}, J\BHBI F.%
, {Schmutz}, W.%
, {Stockman}, Y.%
, {Sch{\"u}hle}, U.%
, {Benmoussa}, A.%
, {Koller}, S.%
\BDBL {}{Rochus}, P.%
\end{APACrefauthors}%
\unskip\
\newblock
\APACrefYearMonthDay{2006}{}{}.
\newblock
{\BBOQ}\APACrefatitle {{LYRA, a solar UV radiometer on Proba2}} {{LYRA, a solar
  UV radiometer on Proba2}}.{\BBCQ}
\newblock
\APACjournalVolNumPages{Advances in Space Research}{37}{}{303-312}.
\newblock
\begin{APACrefDOI} \doi{10.1016/j.asr.2005.10.041} \end{APACrefDOI}
\PrintBackRefs{\CurrentBib}

\bibitem [\protect \citeauthoryear {%
Hunsucker%
\ \BBA {} Hargreaves%
}{%
Hunsucker%
\ \BBA {} Hargreaves%
}{%
{\protect \APACyear {2002}}%
}]{%
Hunsucker_2002}
\APACinsertmetastar {%
Hunsucker_2002}%
\begin{APACrefauthors}%
Hunsucker, R\BPBI D.%
\BCBT {}\ \BBA {} Hargreaves, J\BPBI K.%
\end{APACrefauthors}%
\unskip\
\newblock
\APACrefYear{2002}.
\newblock
\APACrefbtitle {{The High--Latitude Ionosphere and its Effects on Radio
  Propagation}} {{The High--Latitude Ionosphere and its Effects on Radio
  Propagation}}.
\newblock
\APACaddressPublisher{}{Cambridge University Press}.
\PrintBackRefs{\CurrentBib}

\bibitem [\protect \citeauthoryear {%
{ITU-R~P.372-13}%
}{%
{ITU-R~P.372-13}%
}{%
{\protect \APACyear {2016}}%
}]{%
ITU_R}
\APACinsertmetastar {%
ITU_R}%
\begin{APACrefauthors}%
{ITU-R~P.372-13}.%
\end{APACrefauthors}%
\unskip\
\newblock
\APACrefYearMonthDay{2016}{09}{}.
\newblock
\APACrefbtitle {{Recommendation ITU-R P.372-13. Radio noise}.} {{Recommendation
  ITU-R P.372-13. Radio noise}.}
\newblock
\begin{APACrefURL} \url{https://www.itu.int/rec/R-REC-P.372-13-201609-I/en}
  \end{APACrefURL}
\PrintBackRefs{\CurrentBib}

\bibitem [\protect \citeauthoryear {%
Kravtsov%
\ \BBA {} Orlov%
}{%
Kravtsov%
\ \BBA {} Orlov%
}{%
{\protect \APACyear {1983}}%
}]{%
Kravtsov_1983}
\APACinsertmetastar {%
Kravtsov_1983}%
\begin{APACrefauthors}%
Kravtsov, Y.%
\BCBT {}\ \BBA {} Orlov, Y.%
\end{APACrefauthors}%
\unskip\
\newblock
\APACrefYearMonthDay{1983}{}{}.
\newblock
{\BBOQ}\APACrefatitle {Caustics, catastrophes, and wave fields} {Caustics,
  catastrophes, and wave fields}.{\BBCQ}
\newblock
\APACjournalVolNumPages{Soviet Physics Uspekhi}{26}{12}{1038}.
\PrintBackRefs{\CurrentBib}

\bibitem [\protect \citeauthoryear {%
Lawal%
\ \protect \BOthers {.}}{%
Lawal%
\ \protect \BOthers {.}}{%
{\protect \APACyear {2018}}%
}]{%
Lawal_2018}
\APACinsertmetastar {%
Lawal_2018}%
\begin{APACrefauthors}%
Lawal, H\BPBI A.%
, Lester, M.%
, Cowley, S\BPBI W\BPBI H.%
, Milan, S\BPBI E.%
, Yeoman, T\BPBI K.%
, Provan, G.%
\BDBL {}Rabiu, A\BPBI B.%
\end{APACrefauthors}%
\unskip\
\newblock
\APACrefYearMonthDay{2018}{}{}.
\newblock
{\BBOQ}\APACrefatitle {{Understanding the global dynamics of the equatorial
  ionosphere in Africa for space weather capabilities: A science case for
  AfrequaMARN}} {{Understanding the global dynamics of the equatorial
  ionosphere in Africa for space weather capabilities: A science case for
  AfrequaMARN}}.{\BBCQ}
\newblock
\APACjournalVolNumPages{Journal of Atmospheric and Solar-Terrestrial
  Physics}{}{}{}.
\newblock
\begin{APACrefDOI} \doi{10.1016/j.jastp.2018.01.008} \end{APACrefDOI}
\PrintBackRefs{\CurrentBib}

\bibitem [\protect \citeauthoryear {%
Lawson%
\ \BBA {} Hanson%
}{%
Lawson%
\ \BBA {} Hanson%
}{%
{\protect \APACyear {1995}}%
}]{%
Lawson_Hanson_1995}
\APACinsertmetastar {%
Lawson_Hanson_1995}%
\begin{APACrefauthors}%
Lawson, C.%
\BCBT {}\ \BBA {} Hanson, R.%
\end{APACrefauthors}%
\unskip\
\newblock
\APACrefYear{1995}.
\newblock
\APACrefbtitle {Solving Least Squares Problems} {Solving least squares
  problems}.
\newblock
\APACaddressPublisher{}{Society for Industrial and Applied Mathematics}.
\newblock
\begin{APACrefDOI} \doi{10.1137/1.9781611971217} \end{APACrefDOI}
\PrintBackRefs{\CurrentBib}

\bibitem [\protect \citeauthoryear {%
Lemen%
\ \protect \BOthers {.}}{%
Lemen%
\ \protect \BOthers {.}}{%
{\protect \APACyear {2012}}%
}]{%
Lemen_et_al_2012}
\APACinsertmetastar {%
Lemen_et_al_2012}%
\begin{APACrefauthors}%
Lemen, J.%
, Title, A.%
, Akin, D.%
, Boerner, P.%
, Chou, C.%
, Drake, J.%
\BDBL {}Waltham, N.%
\end{APACrefauthors}%
\unskip\
\newblock
\APACrefYearMonthDay{2012}{}{}.
\newblock
{\BBOQ}\APACrefatitle {{The Atmospheric Imaging Assembly (AIA) on the Solar
  Dynamics Observatory (SDO)}} {{The Atmospheric Imaging Assembly (AIA) on the
  Solar Dynamics Observatory (SDO)}}.{\BBCQ}
\newblock
\APACjournalVolNumPages{Solar Physics}{275}{1}{17--40}.
\newblock
\begin{APACrefDOI} \doi{10.1007/s11207-011-9776-8} \end{APACrefDOI}
\PrintBackRefs{\CurrentBib}

\bibitem [\protect \citeauthoryear {%
Liu%
, Hu%
, Liu%
, Wu%
\BCBL {}\ \BBA {} Lester%
}{%
Liu%
\ \protect \BOthers {.}}{%
{\protect \APACyear {2012}}%
}]{%
Liu_2012}
\APACinsertmetastar {%
Liu_2012}%
\begin{APACrefauthors}%
Liu, E\BPBI X.%
, Hu, H\BPBI Q.%
, Liu, R\BPBI Y.%
, Wu, Z\BPBI S.%
\BCBL {}\ \BBA {} Lester, M.%
\end{APACrefauthors}%
\unskip\
\newblock
\APACrefYearMonthDay{2012}{}{}.
\newblock
{\BBOQ}\APACrefatitle {{An adjusted location model for SuperDARN backscatter
  echoes}} {{An adjusted location model for SuperDARN backscatter
  echoes}}.{\BBCQ}
\newblock
\APACjournalVolNumPages{Annales Geophysicae}{30}{12}{1769--1779}.
\newblock
\begin{APACrefDOI} \doi{10.5194/angeo-30-1769-2012} \end{APACrefDOI}
\PrintBackRefs{\CurrentBib}

\bibitem [\protect \citeauthoryear {%
Machol%
\ \BBA {} Viereck%
}{%
Machol%
\ \BBA {} Viereck%
}{%
{\protect \APACyear {2016}}%
}]{%
Machol_2016b}
\APACinsertmetastar {%
Machol_2016b}%
\begin{APACrefauthors}%
Machol, J.%
\BCBT {}\ \BBA {} Viereck, R.%
\end{APACrefauthors}%
\unskip\
\newblock
\APACrefYearMonthDay{2016}{Jun}{10}.
\newblock
\APACrefbtitle {{GOES X--ray Sensor (XRS) Measurements}.} {{GOES X--ray Sensor
  (XRS) Measurements}.}
\newblock
\APAChowpublished
  {https://www.ngdc.noaa.gov/stp/satellite/goes/doc/GOES\_XRS\_readme.pdf}.
\PrintBackRefs{\CurrentBib}

\bibitem [\protect \citeauthoryear {%
Machol%
, Viereck%
\BCBL {}\ \BBA {} Jones%
}{%
Machol%
\ \protect \BOthers {.}}{%
{\protect \APACyear {2016}}%
}]{%
Machol_2016}
\APACinsertmetastar {%
Machol_2016}%
\begin{APACrefauthors}%
Machol, J.%
, Viereck, R.%
\BCBL {}\ \BBA {} Jones, A.%
\end{APACrefauthors}%
\unskip\
\newblock
\APACrefYearMonthDay{2016}{Nov}{08}.
\newblock
\APACrefbtitle {{GOES EUVS Measurements}.} {{GOES EUVS Measurements}.}
\newblock
\APAChowpublished
  {https://www.ngdc.noaa.gov/stp/satellite/goes/doc/GOES\_NOP\_EUV\_readme.pdf}.
\PrintBackRefs{\CurrentBib}

\bibitem [\protect \citeauthoryear {%
Oinats%
, Nishitani%
, Ponomarenko%
, Berngardt%
\BCBL {}\ \BBA {} Ratovsky%
}{%
Oinats%
\ \protect \BOthers {.}}{%
{\protect \APACyear {2016}}%
}]{%
Oinats2016}
\APACinsertmetastar {%
Oinats2016}%
\begin{APACrefauthors}%
Oinats, A.%
, Nishitani, N.%
, Ponomarenko, P.%
, Berngardt, O.%
\BCBL {}\ \BBA {} Ratovsky, K.%
\end{APACrefauthors}%
\unskip\
\newblock
\APACrefYearMonthDay{2016}{}{}.
\newblock
{\BBOQ}\APACrefatitle {{Statistical characteristics of medium--scale traveling
  ionospheric disturbances revealed from the Hokkaido East and Ekaterinburg HF
  radar data}} {{Statistical characteristics of medium--scale traveling
  ionospheric disturbances revealed from the Hokkaido East and Ekaterinburg HF
  radar data}}.{\BBCQ}
\newblock
\APACjournalVolNumPages{{Earth, Planets and Space}}{68}{1}{8}.
\newblock
\begin{APACrefDOI} \doi{10.1186/s40623-016-0390-8} \end{APACrefDOI}
\PrintBackRefs{\CurrentBib}

\bibitem [\protect \citeauthoryear {%
Pederick%
\ \BBA {} Cervera%
}{%
Pederick%
\ \BBA {} Cervera%
}{%
{\protect \APACyear {2014}}%
}]{%
Pederick_2014}
\APACinsertmetastar {%
Pederick_2014}%
\begin{APACrefauthors}%
Pederick, L\BPBI H.%
\BCBT {}\ \BBA {} Cervera, M\BPBI A.%
\end{APACrefauthors}%
\unskip\
\newblock
\APACrefYearMonthDay{2014}{}{}.
\newblock
{\BBOQ}\APACrefatitle {{Semiempirical Model for Ionospheric Absorption based on
  the NRLMSISE--00 atmospheric model}} {{Semiempirical Model for Ionospheric
  Absorption based on the NRLMSISE--00 atmospheric model}}.{\BBCQ}
\newblock
\APACjournalVolNumPages{Radio Science}{49}{2}{81-93}.
\newblock
\begin{APACrefDOI} \doi{10.1002/2013RS005274} \end{APACrefDOI}
\PrintBackRefs{\CurrentBib}

\bibitem [\protect \citeauthoryear {%
Ponomarenko%
, Iserhienrhien%
\BCBL {}\ \BBA {} St.-Maurice%
}{%
Ponomarenko%
\ \protect \BOthers {.}}{%
{\protect \APACyear {2016}}%
}]{%
Ponomarenko_2016}
\APACinsertmetastar {%
Ponomarenko_2016}%
\begin{APACrefauthors}%
Ponomarenko, P.%
, Iserhienrhien, B.%
\BCBL {}\ \BBA {} St.-Maurice, J\BHBI P.%
\end{APACrefauthors}%
\unskip\
\newblock
\APACrefYearMonthDay{2016}{}{}.
\newblock
{\BBOQ}\APACrefatitle {{Morphology and possible origins of near-range oblique
  HF backscatter at high and midlatitudes}} {{Morphology and possible origins
  of near-range oblique HF backscatter at high and midlatitudes}}.{\BBCQ}
\newblock
\APACjournalVolNumPages{Radio Science}{51}{6}{718-730}.
\newblock
\begin{APACrefDOI} \doi{10.1002/2016rs006088} \end{APACrefDOI}
\PrintBackRefs{\CurrentBib}

\bibitem [\protect \citeauthoryear {%
Ponomarenko%
, Nishitani%
, Oinats%
, Tsuya%
\BCBL {}\ \BBA {} St.-Maurice%
}{%
Ponomarenko%
\ \protect \BOthers {.}}{%
{\protect \APACyear {2015}}%
}]{%
Ponomarenko_2015}
\APACinsertmetastar {%
Ponomarenko_2015}%
\begin{APACrefauthors}%
Ponomarenko, P.%
, Nishitani, N.%
, Oinats, A\BPBI V.%
, Tsuya, T.%
\BCBL {}\ \BBA {} St.-Maurice, J\BHBI P.%
\end{APACrefauthors}%
\unskip\
\newblock
\APACrefYearMonthDay{2015}{}{}.
\newblock
{\BBOQ}\APACrefatitle {{Application of ground scatter returns for calibration
  of HF interferometry data}} {{Application of ground scatter returns for
  calibration of HF interferometry data}}.{\BBCQ}
\newblock
\APACjournalVolNumPages{Earth, Planets and Space}{67}{1}{138}.
\newblock
\begin{APACrefDOI} \doi{10.1186/s40623-015-0310-3} \end{APACrefDOI}
\PrintBackRefs{\CurrentBib}

\bibitem [\protect \citeauthoryear {%
Ponomarenko%
, St.-Maurice%
, Hussey%
\BCBL {}\ \BBA {} Koustov%
}{%
Ponomarenko%
\ \protect \BOthers {.}}{%
{\protect \APACyear {2010}}%
}]{%
Ponomarenko_2010}
\APACinsertmetastar {%
Ponomarenko_2010}%
\begin{APACrefauthors}%
Ponomarenko, P.%
, St.-Maurice, J\BHBI P.%
, Hussey, G.%
\BCBL {}\ \BBA {} Koustov, A.%
\end{APACrefauthors}%
\unskip\
\newblock
\APACrefYearMonthDay{2010}{}{}.
\newblock
{\BBOQ}\APACrefatitle {{HF ground scatter from the polar cap: Ionospheric
  propagation and ground surface effects}} {{HF ground scatter from the polar
  cap: Ionospheric propagation and ground surface effects}}.{\BBCQ}
\newblock
\APACjournalVolNumPages{J. Geophys. Res}{115}{}{10310}.
\newblock
\begin{APACrefDOI} \doi{10.1029/2010JA015828} \end{APACrefDOI}
\PrintBackRefs{\CurrentBib}

\bibitem [\protect \citeauthoryear {%
Ponomarenko%
\ \BBA {} Waters%
}{%
Ponomarenko%
\ \BBA {} Waters%
}{%
{\protect \APACyear {2006}}%
}]{%
Ponomarenko_2006}
\APACinsertmetastar {%
Ponomarenko_2006}%
\begin{APACrefauthors}%
Ponomarenko, P.%
\BCBT {}\ \BBA {} Waters, C.%
\end{APACrefauthors}%
\unskip\
\newblock
\APACrefYearMonthDay{2006}{}{}.
\newblock
{\BBOQ}\APACrefatitle {{Spectral width of SuperDARN echoes: measurement, use
  and physical interpretation}} {{Spectral width of SuperDARN echoes:
  measurement, use and physical interpretation}}.{\BBCQ}
\newblock
\APACjournalVolNumPages{Annales Geophysicae}{24}{1}{115--128}.
\newblock
\begin{APACrefDOI} \doi{10.5194/angeo-24-115-2006} \end{APACrefDOI}
\PrintBackRefs{\CurrentBib}

\bibitem [\protect \citeauthoryear {%
Ribeiro%
\ \protect \BOthers {.}}{%
Ribeiro%
\ \protect \BOthers {.}}{%
{\protect \APACyear {2012}}%
}]{%
Ribeiro_2012}
\APACinsertmetastar {%
Ribeiro_2012}%
\begin{APACrefauthors}%
Ribeiro, A\BPBI J.%
, Ruohoniemi, J.%
, Baker, J.%
, Clausen, L\BPBI B\BPBI N.%
, Greenwald, R.%
\BCBL {}\ \BBA {} Lester, M.%
\end{APACrefauthors}%
\unskip\
\newblock
\APACrefYearMonthDay{2012}{}{}.
\newblock
{\BBOQ}\APACrefatitle {{A survey of plasma irregularities as seen by the
  midlatitude Blackstone SuperDARN radar}} {{A survey of plasma irregularities
  as seen by the midlatitude Blackstone SuperDARN radar}}.{\BBCQ}
\newblock
\APACjournalVolNumPages{Journal of Geophysical Research: Space
  Physics}{117}{A2}{A02311}.
\newblock
\begin{APACrefDOI} \doi{10.1029/2011ja017207} \end{APACrefDOI}
\PrintBackRefs{\CurrentBib}

\bibitem [\protect \citeauthoryear {%
Ribeiro%
\ \protect \BOthers {.}}{%
Ribeiro%
\ \protect \BOthers {.}}{%
{\protect \APACyear {2011}}%
}]{%
Ribeiro_2011}
\APACinsertmetastar {%
Ribeiro_2011}%
\begin{APACrefauthors}%
Ribeiro, A\BPBI J.%
, Ruohoniemi, J.%
, Baker, J.%
, Clausen, S.%
, de Larquier, S.%
\BCBL {}\ \BBA {} Greenwald, R.%
\end{APACrefauthors}%
\unskip\
\newblock
\APACrefYearMonthDay{2011}{}{}.
\newblock
{\BBOQ}\APACrefatitle {{A new approach for identifying ionospheric backscatter
  in midlatitude SuperDARN HF radar observations}} {{A new approach for
  identifying ionospheric backscatter in midlatitude SuperDARN HF radar
  observations}}.{\BBCQ}
\newblock
\APACjournalVolNumPages{Radio Sci.}{46}{}{RS4011}.
\newblock
\begin{APACrefDOI} \doi{10.1029/2011RS004676} \end{APACrefDOI}
\PrintBackRefs{\CurrentBib}

\bibitem [\protect \citeauthoryear {%
Rogers%
\ \BBA {} Honary%
}{%
Rogers%
\ \BBA {} Honary%
}{%
{\protect \APACyear {2015}}%
}]{%
Rogers_2015}
\APACinsertmetastar {%
Rogers_2015}%
\begin{APACrefauthors}%
Rogers, N\BPBI C.%
\BCBT {}\ \BBA {} Honary, F.%
\end{APACrefauthors}%
\unskip\
\newblock
\APACrefYearMonthDay{2015}{}{}.
\newblock
{\BBOQ}\APACrefatitle {{Assimilation of real--time riometer measurements into
  models of 30~MHz polar cap absorption}} {{Assimilation of real--time riometer
  measurements into models of 30~MHz polar cap absorption}}.{\BBCQ}
\newblock
\APACjournalVolNumPages{J. Space Weather Space Clim.}{5}{}{A8}.
\newblock
\begin{APACrefDOI} \doi{10.1051/swsc/2015009} \end{APACrefDOI}
\PrintBackRefs{\CurrentBib}

\bibitem [\protect \citeauthoryear {%
Sauer%
\ \BBA {} Wilkinson%
}{%
Sauer%
\ \BBA {} Wilkinson%
}{%
{\protect \APACyear {2008}}%
}]{%
Sauer_2008}
\APACinsertmetastar {%
Sauer_2008}%
\begin{APACrefauthors}%
Sauer, H\BPBI H.%
\BCBT {}\ \BBA {} Wilkinson, D\BPBI C.%
\end{APACrefauthors}%
\unskip\
\newblock
\APACrefYearMonthDay{2008}{}{}.
\newblock
{\BBOQ}\APACrefatitle {{Global mapping of ionospheric HF/VHF radio wave
  absorption due to solar energetic protons}} {{Global mapping of ionospheric
  HF/VHF radio wave absorption due to solar energetic protons}}.{\BBCQ}
\newblock
\APACjournalVolNumPages{Space Weather}{6}{12}{}.
\newblock
\APACrefnote{S12002}
\newblock
\begin{APACrefDOI} \doi{10.1029/2008sw000399} \end{APACrefDOI}
\PrintBackRefs{\CurrentBib}

\bibitem [\protect \citeauthoryear {%
Schumer%
}{%
Schumer%
}{%
{\protect \APACyear {2010}}%
}]{%
Schumer2010}
\APACinsertmetastar {%
Schumer2010}%
\begin{APACrefauthors}%
Schumer, E\BPBI A.%
\end{APACrefauthors}%
\unskip\
\newblock
\APACrefYear{2010}.
\unskip\
\newblock
\APACrefbtitle {{Improved modeling of midlatitude D--region ionospheric
  absorption of high frequency radio signals during solar X--ray flares}}
  {{Improved modeling of midlatitude D--region ionospheric absorption of high
  frequency radio signals during solar X--ray flares}}\ \APACtypeAddressSchool
  {Ph.D. dissertation}{}{}.
\unskip\
\newblock
\APACaddressSchool {}{Air Force Institute of Technology}.
\PrintBackRefs{\CurrentBib}

\bibitem [\protect \citeauthoryear {%
Shearman%
}{%
Shearman%
}{%
{\protect \APACyear {1956}}%
}]{%
Shearman_1956}
\APACinsertmetastar {%
Shearman_1956}%
\begin{APACrefauthors}%
Shearman, E\BPBI D\BPBI R.%
\end{APACrefauthors}%
\unskip\
\newblock
\APACrefYearMonthDay{1956}{}{}.
\newblock
{\BBOQ}\APACrefatitle {A study of ionospheric propagation by means of ground
  back--scatter} {A study of ionospheric propagation by means of ground
  back--scatter}.{\BBCQ}
\newblock
\APACjournalVolNumPages{Proceedings of the IEE - Part B: Radio and Electronic
  Engineering}{103}{8}{203-209}.
\newblock
\begin{APACrefDOI} \doi{10.1049/pi-b-1.1956.0145} \end{APACrefDOI}
\PrintBackRefs{\CurrentBib}

\bibitem [\protect \citeauthoryear {%
Shepherd%
}{%
Shepherd%
}{%
{\protect \APACyear {2017}}%
}]{%
Shepherd_2017}
\APACinsertmetastar {%
Shepherd_2017}%
\begin{APACrefauthors}%
Shepherd, S\BPBI G.%
\end{APACrefauthors}%
\unskip\
\newblock
\APACrefYearMonthDay{2017}{}{}.
\newblock
{\BBOQ}\APACrefatitle {{Elevation angle determination for SuperDARN HF radar
  layouts}} {{Elevation angle determination for SuperDARN HF radar
  layouts}}.{\BBCQ}
\newblock
\APACjournalVolNumPages{Radio Science}{52}{8}{938-950}.
\newblock
\APACrefnote{2017RS006348}
\newblock
\begin{APACrefDOI} \doi{10.1002/2017rs006348} \end{APACrefDOI}
\PrintBackRefs{\CurrentBib}

\bibitem [\protect \citeauthoryear {%
Simon%
}{%
Simon%
}{%
{\protect \APACyear {2013}}%
}]{%
simon2013}
\APACinsertmetastar {%
simon2013}%
\begin{APACrefauthors}%
Simon, D.%
\end{APACrefauthors}%
\unskip\
\newblock
\APACrefYear{2013}.
\newblock
\APACrefbtitle {{Evolutionary Optimization Algorithms}} {{Evolutionary
  Optimization Algorithms}}.
\newblock
\APACaddressPublisher{}{Wiley}.
\PrintBackRefs{\CurrentBib}

\bibitem [\protect \citeauthoryear {%
Squibb%
\ \protect \BOthers {.}}{%
Squibb%
\ \protect \BOthers {.}}{%
{\protect \APACyear {2015}}%
}]{%
Squibb_2015}
\APACinsertmetastar {%
Squibb_2015}%
\begin{APACrefauthors}%
Squibb, C\BPBI O.%
, Frissell, N\BPBI A.%
, Ruohoniemi, J\BPBI M.%
, Baker, J\BPBI B\BPBI H.%
, Fiori, R.%
\BCBL {}\ \BBA {} Moses, M\BPBI L.%
\end{APACrefauthors}%
\unskip\
\newblock
\APACrefYearMonthDay{2015}{}{}.
\newblock
{\BBOQ}\APACrefatitle {{Dayside Ionospheric Response to X--Class Solar Flare
  Events Observed with Reverse Beacon Network High Frequency Communication
  Links}} {{Dayside Ionospheric Response to X--Class Solar Flare Events
  Observed with Reverse Beacon Network High Frequency Communication
  Links}}.{\BBCQ}
\newblock
\BIn{} \APACrefbtitle {Virginia Tech REU Symposium -- Poster Presentation.}
  {Virginia tech reu symposium -- poster presentation.}
\newblock
\APACaddressPublisher{Blacksburg, VA}{Virginia Tech REU Program}.
\PrintBackRefs{\CurrentBib}

\bibitem [\protect \citeauthoryear {%
Stocker%
, Arnold%
\BCBL {}\ \BBA {} Jones%
}{%
Stocker%
\ \protect \BOthers {.}}{%
{\protect \APACyear {2000}}%
}]{%
Stocker_2000}
\APACinsertmetastar {%
Stocker_2000}%
\begin{APACrefauthors}%
Stocker, A\BPBI J.%
, Arnold, N\BPBI F.%
\BCBL {}\ \BBA {} Jones, T\BPBI B.%
\end{APACrefauthors}%
\unskip\
\newblock
\APACrefYearMonthDay{2000}{}{}.
\newblock
{\BBOQ}\APACrefatitle {{The synthesis of travelling ionospheric disturbance
  (TID) signatures in HF radar observations using ray tracing}} {{The synthesis
  of travelling ionospheric disturbance (TID) signatures in HF radar
  observations using ray tracing}}.{\BBCQ}
\newblock
\APACjournalVolNumPages{Annales Geophysicae}{18}{1}{56--64}.
\newblock
\begin{APACrefDOI} \doi{10.1007/s00585-000-0056-4} \end{APACrefDOI}
\PrintBackRefs{\CurrentBib}

\bibitem [\protect \citeauthoryear {%
Tinin%
}{%
Tinin%
}{%
{\protect \APACyear {1983}}%
}]{%
Tinin1983}
\APACinsertmetastar {%
Tinin1983}%
\begin{APACrefauthors}%
Tinin, M\BPBI V.%
\end{APACrefauthors}%
\unskip\
\newblock
\APACrefYearMonthDay{1983}{}{}.
\newblock
{\BBOQ}\APACrefatitle {Propagation of waves in a medium with large--scale
  random inhomogeneities} {Propagation of waves in a medium with large--scale
  random inhomogeneities}.{\BBCQ}
\newblock
\APACjournalVolNumPages{Radiophysics and Quantum Electronics}{26}{1}{29--36}.
\newblock
\begin{APACrefDOI} \doi{10.1007/BF01038771} \end{APACrefDOI}
\PrintBackRefs{\CurrentBib}

\bibitem [\protect \citeauthoryear {%
Tobiska%
, Bouwer%
\BCBL {}\ \BBA {} Bowman%
}{%
Tobiska%
\ \protect \BOthers {.}}{%
{\protect \APACyear {2008}}%
}]{%
TOBISKA2008803}
\APACinsertmetastar {%
TOBISKA2008803}%
\begin{APACrefauthors}%
Tobiska, K\BPBI W.%
, Bouwer, D\BPBI S.%
\BCBL {}\ \BBA {} Bowman, B\BPBI R.%
\end{APACrefauthors}%
\unskip\
\newblock
\APACrefYearMonthDay{2008}{}{}.
\newblock
{\BBOQ}\APACrefatitle {The development of new solar indices for use in
  thermospheric density modeling} {The development of new solar indices for use
  in thermospheric density modeling}.{\BBCQ}
\newblock
\APACjournalVolNumPages{Journal of Atmospheric and Solar-Terrestrial
  Physics}{70}{5}{803 - 819}.
\newblock
\begin{APACrefDOI} \doi{10.1016/j.jastp.2007.11.001} \end{APACrefDOI}
\PrintBackRefs{\CurrentBib}

\bibitem [\protect \citeauthoryear {%
Uryadov%
, Vertogradov%
, Sklyarevsky%
\BCBL {}\ \BBA {} Vybornov%
}{%
Uryadov%
\ \protect \BOthers {.}}{%
{\protect \APACyear {2018}}%
}]{%
Uryadov_2018}
\APACinsertmetastar {%
Uryadov_2018}%
\begin{APACrefauthors}%
Uryadov, V\BPBI P.%
, Vertogradov, G.%
, Sklyarevsky, M\BPBI S.%
\BCBL {}\ \BBA {} Vybornov, F\BPBI I.%
\end{APACrefauthors}%
\unskip\
\newblock
\APACrefYearMonthDay{2018}{}{}.
\newblock
{\BBOQ}\APACrefatitle {{Positioning of Ionospheric Irregularities and the
  Earth's Surface Roughness Using an Over-the-Horizon HF Radar}} {{Positioning
  of Ionospheric Irregularities and the Earth's Surface Roughness Using an
  Over-the-Horizon HF Radar}}.{\BBCQ}
\newblock
\APACjournalVolNumPages{Radiophysics and Quantum Electronics}{}{}{}.
\newblock
\begin{APACrefDOI} \doi{10.1007/s11141-018-9838-y} \end{APACrefDOI}
\PrintBackRefs{\CurrentBib}

\bibitem [\protect \citeauthoryear {%
Warrington%
\ \protect \BOthers {.}}{%
Warrington%
\ \protect \BOthers {.}}{%
{\protect \APACyear {2016}}%
}]{%
Warrington_2016}
\APACinsertmetastar {%
Warrington_2016}%
\begin{APACrefauthors}%
Warrington, E.%
, Stocker, A\BPBI J.%
, Siddle, D\BPBI R.%
, Hallam, J.%
, Al-Behadili, H\BPBI A\BPBI H.%
, Zaalov, N\BPBI Y.%
\BDBL {}Danskin, D.%
\end{APACrefauthors}%
\unskip\
\newblock
\APACrefYearMonthDay{2016}{}{}.
\newblock
{\BBOQ}\APACrefatitle {{Near real--time input to a propagation model for
  nowcasting of HF communications with aircraft on polar routes}} {{Near
  real--time input to a propagation model for nowcasting of HF communications
  with aircraft on polar routes}}.{\BBCQ}
\newblock
\APACjournalVolNumPages{Radio Science}{}{}{1048--1059}.
\newblock
\APACrefnote{2015RS005880}
\newblock
\begin{APACrefDOI} \doi{10.1002/2015rs005880} \end{APACrefDOI}
\PrintBackRefs{\CurrentBib}

\bibitem [\protect \citeauthoryear {%
Watanabe%
\ \BBA {} Nishitani%
}{%
Watanabe%
\ \BBA {} Nishitani%
}{%
{\protect \APACyear {2013}}%
}]{%
Watanabe_2014}
\APACinsertmetastar {%
Watanabe_2014}%
\begin{APACrefauthors}%
Watanabe, D.%
\BCBT {}\ \BBA {} Nishitani, N.%
\end{APACrefauthors}%
\unskip\
\newblock
\APACrefYearMonthDay{2013}{}{}.
\newblock
{\BBOQ}\APACrefatitle {{Study of ionospheric disturbances during solar flare
  events using the SuperDARN Hokkaido radar}} {{Study of ionospheric
  disturbances during solar flare events using the SuperDARN Hokkaido
  radar}}.{\BBCQ}
\newblock
\APACjournalVolNumPages{Advances in Polar Science}{24}{1}{12--18}.
\newblock
\begin{APACrefDOI} \doi{10.3724/sp.j.1085.2013.00012} \end{APACrefDOI}
\PrintBackRefs{\CurrentBib}

\bibitem [\protect \citeauthoryear {%
Weisstein%
}{%
Weisstein%
}{%
{\protect \APACyear {{\protect \bibnodate {}}}}%
}]{%
Weisstein_2018}
\APACinsertmetastar {%
Weisstein_2018}%
\begin{APACrefauthors}%
Weisstein, E.%
\end{APACrefauthors}%
\unskip\
\newblock
\APACrefYearMonthDay{{\protect \bibnodate {}}}{}{}.
\newblock
\APACrefbtitle {{Parabolic Cylinder Function}.} {{Parabolic Cylinder
  Function}.}
\newblock
\APAChowpublished
  {http://mathworld.wolfram.com/ParabolicCylinderFunction.html}.
\PrintBackRefs{\CurrentBib}

\bibitem [\protect \citeauthoryear {%
Yukimatu%
\ \BBA {} Tsutsumi%
}{%
Yukimatu%
\ \BBA {} Tsutsumi%
}{%
{\protect \APACyear {2002}}%
}]{%
Yukimatu_2002}
\APACinsertmetastar {%
Yukimatu_2002}%
\begin{APACrefauthors}%
Yukimatu, A\BPBI S.%
\BCBT {}\ \BBA {} Tsutsumi, M.%
\end{APACrefauthors}%
\unskip\
\newblock
\APACrefYearMonthDay{2002}{}{}.
\newblock
{\BBOQ}\APACrefatitle {{A new SuperDARN meteor wind measurement: Raw time
  series analysis method and its application to mesopause region dynamics}} {{A
  new SuperDARN meteor wind measurement: Raw time series analysis method and
  its application to mesopause region dynamics}}.{\BBCQ}
\newblock
\APACjournalVolNumPages{Geophysical Research Letters}{29}{20}{42-1-42-4}.
\newblock
\begin{APACrefDOI} \doi{10.1029/2002GL015210} \end{APACrefDOI}
\PrintBackRefs{\CurrentBib}

\end{thebibliography}

\end{document}